%% file: bershady.tex
\documentstyle[emulateaj]{article}

\newcommand\etal{{\it et~al.\ }}

\begin{document}

\submitted{To Appear in AJ, June 2000}

\title{Structural and Photometric Classification of Galaxies -- 
I. Calibration Based on a Nearby Galaxy Sample}

\author{Matthew A. Bershady}

\affil{Department of Astronomy, University of Wisconsin, 475 N Charter
Street, Madison, WI 53706; mab@astro.wisc.edu}

\author{Anna Jangren}

\affil{Department of Astronomy and Astrophysics, Pennsylvania State
University, 525 Davey Lab, University Park, PA 16802;
jangren@astro.psu.edu}

\author{Christopher J. Conselice}

\affil{Department of Astronomy, University of Wisconsin, 475 N Charter
Street, Madison, WI 53706; chris@astro.wisc.edu}

%
%
%


\begin{abstract}

In this paper we define an observationally robust, multi-parameter
space for the classification of nearby and distant galaxies. The
parameters include luminosity, color, and the image-structure
parameters: size, image concentration, asymmetry, and surface
brightness. Based on an initial calibration of this parameter space
using the ``normal'' Hubble-types surveyed by Frei \etal (1996), we
find that only a subset of the parameters provide useful
classification boundaries for this sample.  Interestingly, this subset
does not include distance-dependent scale parameters, such as size or
luminosity. The essential ingredient is the combination of a spectral
index (e.g., color) with parameters of image structure and scale:
concentration, asymmetry, and surface-brightness.  We refer to the
image structure parameters (concentration and asymmetry) as indices of
``form.'' We define a preliminary classification based on spectral
index, form, and surface-brightness (a scale) that successfully
separates normal galaxies into three classes. We intentionally
identify these classes with the familiar labels of Early,
Intermediate, and Late. This classification, or others based on the
above four parameters can be used reliably to define comparable
samples over a broad range in redshift. The size and luminosity
distribution of such samples will not be biased by this selection
process except through astrophysical correlations between spectral
index, form, and surface-brightness.

\end{abstract}

\keywords{galaxies: classification --- galaxies: morphology ---
galaxies: colors}
              
\section{Introduction}

It is now well-established that a large fraction of galaxies
discovered at intermediate and high redshift have unusual
morphologies, and thus cannot be classified in terms of the nominal
Hubble-Sandage system (Driver \etal 1995, 1998; Abraham \etal 1996a,
1996b). The Hubble classification scheme is also difficult to apply
to many local galaxies, dubbed `peculiar,' or any galaxies
imaged at low signal-to-noise ($S/N$), or apparently small size
(relative to the point-spread function). The Hubble-Sandage
classification system was predicated on the study of nearby,
``normal'' galaxies -- luminous and relatively quiescent
objects(\cite{san61}, \cite{san87}, \cite{san93}). While the
classification system developed by de~Vaucouleurs \etal (1976) makes
an attempt to push the framework to ``later'' types, it still suffers
from the above shortcomings. Fundamentally, these traditional
classification schemes are based on the concept of `pigeon-holing'
galaxies based on a reference-set, or archetypes. These archetypes are
selected from samples in the local universe, and are preferentially
axisymmetric systems.  Since our local census is undoubtedly
incomplete, and, since galaxies evolve, such reference sets by their
very definition are incomplete. Thus it is not surprising that these
systems are of marginal utility in the study of dwarf galaxies,
interacting galaxies, or galaxies at high redshift.

An alternative classification scheme could be based on quantitative
indices, the inter-relation of which is not predetermined by a finite
reference set. This would permit galaxies to be classified, for
example, in different stages of their evolution; albeit the
classification would be different but the basis set of indices would
be the same. The goal of this paper is to define such a set of
indices that can be used as quantitative, objective classifiers of
galaxies (i) over a wide range in redshift, and (ii) for wide range of
galaxy types. In particular, we desire classifiers that are well
suited to typing both ``normal'' galaxies and the compact galaxies
that are the focus of a companion study (Jangren \etal 2000;
hereafter, paper II). We anticipate that such a classification scheme
is both necessary and enabling for the exploration of the physical
mechanisms driving galaxy evolution (Bershady 1999).

What are the desired characteristics of classification parameters?
They should be physically interesting (closely related to underlying
physical properties of galaxies), model-independent, and measurable
for all galaxy types. It also should be possible to accurately
determine the parameters chosen for a wide range of image resolution 
and signal-to-noise ratios.

From Hubble's classification {\it a posteriori} we have learned that a
strong correlation exists between galaxy spectral type and apparent
morphological features -- at least for the galaxy types which fit well
within his scheme. This correlation -- noted by Hubble as early
as 1936 (Hubble 1936) -- can loosely be termed a `color-morphology'
relation, although the correlation is not necessarily limited to
broad-band color. This is a triumph of Hubble's classification
explicitly because it is not part {\it of} the classification.
Furthermore, the correlation yields clues about the physical
connection of the present matter distribution and the star-formation
histories in galaxies. But while morphology (or `form') and spectral
type are correlated, there is also significant dispersion in this
correlation. Some of the more notable deviations from the nominal
color-morphology relation are found in the plethora of forms for
spectrally `late' type galaxies, the presence of `E+A' galaxies
(Dressler \& Gunn 1993), and the compact, luminous, blue, emission-line
galaxies studied in paper II (as we shall show). This points to the
importance of form and spectral type as key, yet independent axes of a
revised classification system.

However, the only example of such a revised classification system is
that of Morgan (1958, 1959), where central light concentration is used
as the primary classification parameter. Morgan was motivated by the
fact that (i) a salient criterion used in classifying galaxies in the
Hubble-Sandage system is the degree of central concentration of light;
(ii) there was a significant dispersion in spectral type and Hubble
type (\cite{hum56}); and (iii) spectral type appeared to correlate
more strongly with light concentration. In this way, Morgan hoped to
wed the classification of stellar populations to the classification of
galaxies. Nonetheless, he was compelled to introduce a secondary
parameter, i.e. the `Form Family,' because there was still a
dispersion of morphological forms within each of his spectral types.
Today, one should be able to improve upon Morgan's scheme by
introducing quantitative measures of image concentration and other
indices of form, and by independently assessing the spectral type via
colors or spectra.

A number of subsequent attempts have been made to construct
quantitative classification system that could replace or modify the
current Hubble scheme. Yet these schemes are generally based purely
either on photometric form (e.g. \cite{elm82}; \cite{oka84};
\cite{wat85}; \cite{doi93}; \cite{abr94}; \cite{ode95}; \cite{han95})
or spectral type (e.g. \cite{ber95}; \cite{conn95}; \cite{zar95};
\cite{fol96}; \cite{bro98}; \cite{ron99}). In essence, they have
relied implicitly on an assumed correlation between galaxy spectral
type and apparent morphology. Related attempts have been made to use
artificial neural networks to reproduce the Hubble scheme in an
objective way (e.g. \cite{bur92}; \cite{sto92}; \cite{spie92};
\cite{ser93}; \cite{nai95}; \cite{ode95}; \cite{ode96}). Yet these go
no further in differentiating between spectral type and form. Only in
Whitmore's (1984) scheme are spectral and structural parameters
combined, i.e., $B-H$ color, size, and bulge-to-total ratio are used
to define two principal classification axes of {\it scale} and {\it
form}. But again, the correlation(s) between galaxy spectral type,
scale and form are not explicit.

Here we attempt to expand on Morgan's program by fully quantifying the
classification of form via image concentration and several other
structural parameters, and explicitly using color as an indicator of
spectral type. In this study we choose to use only a single color
($B-V$), but we anticipate that a more desirable, future development
would be to include broad-wavelength coverage, multi-color data and
spectroscopic line-indices. Spectroscopic line-indices would be
required, for example, to identify E+A galaxies. While such galaxies
are not the focus of the present work, a comprehensive classification
scheme should be able to isolate these systems and determine the range
of their morphology ({\it cf.} Dressler \& Gunn 1992, Couch \etal
1994, and Wirth \etal 1994). Nonetheless, broad-band colors are a
cost-effective way to characterize the spectral continuum ({\it cf.}
Bershady, 1995, and Connolly \etal 1995). Of more direct relevance to
the study at hand, a future elaboration of including $U-V$ and $V-K$
would enhance the ability to distinguish between spectral types
particularly for galaxies with extremely blue, optical colors
(e.g. Aaronson 1978, Bershady 1995).

We have also chosen to quantify form and scale via non-parametric
measures, such as luminosity, half-light size and surface-brightness,
asymmetry, and image concentration. An alternative, model-dependent
approach is to decompose a galaxy's light profile into a disk and
bulge. The traditional one-dimensional decompositions are fraught with
technical problems such that decompositions can only be achieved
reliably for about half of all disk galaxies (\cite{ken85}). The newer
two-dimensional decomposition techniques are superior (e.g.
\cite{jong96b}), and have been shown to successfully reproduce
observed light profiles for faint galaxies (e.g. \cite{sim99}).
Indeed, one can argue that two-dimensional model fitting to imaging
data is optimum in terms of using the available information, and for
minimizing random error. At high $S/N$ and high angular resolution,
however, even the most ``normal'' galaxies exhibit peculiarities (as
discussed in more detail in \S 3.3.2) such that simple bulge-plus-disk
models cannot reproduce these frequently observed peculiarities in
light distributions with high fidelity. The situation worsens for
``peculiar'' galaxies. For this reason we have some concerns about the
uniqueness of the observationally derived model parameters, and hence
their interpretation. We anticipate future developments which use the
models and non-parametric measurements in a hybrid scheme optimal for
characterizing galaxy light distributions both in terms of random and
systematic errors.

It is worth noting again that bright galaxy samples are notorious for
missing or under-representing certain galaxy types -- particularly
dwarfs and low-surface-brightness galaxies. The samples used here are
no exception. While this was one of our complaints about the classical
Hubble scheme, there are two key differences with our approach: (i)
the classification parameters we develop are objective; and (ii) these
parameters do not assume the presence of basic axi-symmetry,
disk-plus-bulge structure, or spiral patterns which underly the Hubble
scheme. As we will show, the galaxies examined here are sufficiently
diverse to establish the {\it parameter space} for a comprehensive
classification scheme, although not the comprehensive classification
itself. By developing an initial classification of these galaxies,
however, we intend to use it as a foil against which we can begin to
compare the classification of more distant samples: How are the
classifications different? Do the nearby and distant samples occupy
the same regions of parameter space? If not, do the differences
represent continuous extensions of these parameters, or are they
physically disjoint? These are the types of questions one can address
given the limitations of current local samples.  Note that we must
stop short of identifying differences as ``new,'' epoch-specific
classes of galaxies. Without a complete census of both the nearby and
distant universe, it is not possible to establish whether there are
different ``classes'' of galaxies at different redshifts; apparent
differences could simply be artifacts of the presently limited
samples. With such a complete census, in the future we may hope to
address deeper issue of how the comoving space-densities of different
classes evolve.

Towards the goal of establishing a comprehensive classification scheme
of utility to distant galaxy studies, in this paper we assemble a
robust set of non-parametric, photometric and structural properties
for a range of nearby, lumimous galaxies. We define a multivariate,
photometric parameter space that forms an initial classification
scheme for these galaxies. This classification can be used reliably to
identify comparable samples in other surveys and at higher redshift.
In the accompanying paper (II) we measure these properties for
compact, luminous emission-line galaxies at intermediate redshift,
compare them to the ``normal,'' nearby galaxies studied here, and
demonstrate that our classification parameter space distinguishes
between these two samples. We discuss the implications for the
evolution of this intermediate-redshift sample therein.  In future
papers in this series we intend to extend our analysis (a) to more
representative samples of the local volume that include dwarf and
emission-line galaxies (e.g., the University of Michigan Objective
Prism Survey (\cite{sal89}); (b) to more comprehensive samples of
distant galaxies, e.g magnitude-limited samples from the Hubble Deep
Field; and (c) to studies of the morphological evolution of these
distant samples. The classification scheme which we propose here is
intended as a framework for these future studies.

The data sets are presented in \S 2; the analysis is described in \S
3. The results are presented in \S 4, and summarized in \S
5. Throughout this paper we adopt $H_0$ = 50 km s$^{-1}$ Mpc$^{-1}$,
$q_0 = 0.1$, $\Lambda = 0$.

\section{Nearby galaxy samples} 

As a primary reference sample, 101 of the 113 local Hubble-type
galaxies from the catalog of Frei \etal (1996) were analyzed.  This
sample will define what we mean by ``normal'' galaxies in this paper.
This catalog is the only digital, multi-band, sample publicly
available that is reasonably comprehensive; it consists of
ground-based CCD images of bright galaxies, all apparently large (most
have diameters of 4\arcmin~-~6\arcmin) and well resolved. As a result,
the sample contains mostly luminous and physically large galaxies: out
of the 101 objects we used in our analysis, only seven have $L <
0.1\,L^*$. We excluded 12 objects whose apparent sizes were larger
than the CCD field of view (thus their image structure parameters
could not be well estimated). Two of the excluded objects are
early-type galaxies (E--S0), seven are intermediate (Sa--Sb), and
three are late-type (Sc--Irr).\footnote{The excluded objects are: NGC
2403, 2683, 3031, 3079, 3351, 3623, 4406, 4472, 4594, 4826, 5746, and
6503.} The majority of the remaining sample are spirals and S0
galaxies. Frei \etal have removed foreground stars from the images of
the nearby galaxies, in a few cases leaving visible ``scars;'' except
in the case of NGC 5792, these residuals did not cause noticeable
problems when determining the structural parameters (\S 3.3).

In several instances in the present analysis we reference the sample
of Kent (1984, 1985), which is composed of 53 nearby, luminous and
physically large galaxies similar to the Frei \etal sample. We find
Kent's sample useful for comparison of both photometric and structural
parameters. We also reference the sample of 196 normal (non-active)
Markarian galaxies studied by Huchra (1977a). Relevant characteristics
of the above three samples are summarized in Table 1, including an
enumeration of the effective filter systems used in each
study. Further details on these photometric systems are found in the
studies listed in the Table and references therein.

\subsection{Comparison of reference samples to emission-line galaxy samples}

Both the Frei \etal and Kent samples are under-representative of dwarf
galaxies, and contain neither HII galaxies nor low surface-brightness
galaxies. The latter objects have been shown to make up a significant
fraction of the local galaxy population (de Jong 1995, 1996a). Clearly
our reference samples do not constitute a representative template of
the local population. Here we estimate where these samples may be
particularly un-representative with an eye towards the study of faint
galaxy samples in future papers. In Figures 1 and 2 we compare the
Frei \etal samples photometric properties of color and luminosity to
(i) the normal Markarian galaxies (Huchra 1977a), (ii) dwarf
spheroidals (as described in the following section), and (iii) the
intermediate redshift samples presented in paper II.

Since the Markarian galaxies were selected from objective prism plates
based on their strong UV continua, the sample is biased toward bluer
colors than the Frei \etal galaxies and is thus likely more
representative of star-forming galaxies. Huchra's sample contains
fainter galaxies that extend the magnitude range down to $M_B \sim
-14$ and the color-color locus blue-ward of $B-V = 0.4$.

The intermediate-redshift galaxies, also selected in part due to their
blue color (see paper II), have blue luminosities comparable to the
brighter half of the Frei \etal sample, but with bluer colors. This
places most of them in a distinct region of the color-luminosity plot
from the Frei \etal sample. In contrast, the distribution of the
Markarian galaxies extends into the region occupied by the
intermediate-redshift objects. In the color-color diagram, again the
intermediate-redshift galaxies largely overlap with the Markarian
sample in the region corresponding to extreme blue colors {\it not}
occupied by the Frei \etal or Kent samples.

In short, the Frei \etal sample is spectro-photometrically disjoint
from extreme samples of blue, star-forming galaxies at
intermediate-redshift (e.g., paper II), even though both contain
intrinsically luminous and moderate-to-high surface-brightness
systems. Yet clearly there ar local examples (e.g., from Markarian)
which are as blue and luminous as these intermediate-redshift,
star-forming galaxies. These sources are simply missing from the Frei
\etal sample.  The comparison of the global properties of the
intermediate-redshift, compact, star-forming galaxies in paper II to
those of local galaxies from Frei \etal (here) is then an initial step
in mapping the range of galaxy types at any redshift. Further
investigation of the nature and evolution of these types of extreme,
star-forming systems will be greatly facilitated by future work
quantifying the image structure of local counterparts \bv $< 0.4$ and
$M_B<-19$.

\subsection{Comparison of reference samples to dwarf spheroidals}

We have made some attempt, where possible, to access the photometric
and structural properties of other key dwarf populations. We
schematically indicate the locus of dwarf ellipticals/spheroidals in
Figures 1 and 2 using data from \cite{cald83}, \cite{bing91}, and
\cite{bing98}. The dwarf spheroidals occupy a virtually unpopulated
region of the color-luminosity diagram at relatively red colors and
low luminosity. The absence of such objects from most surveys is
attributed typically to a selection bias since these sources are at
low surface-brightness. It is interesting to note that in the
color-color diagram the dwarf spheroidals occupy a region over-lapping
with the early-to-intermediate type spirals. Hence the integrated
broad-band light of these systems are unusual compared to our
reference samples only with respect to their luminosity.  We refer to
the dwarf spheroidal properties extensively in future papers where we
also explore their image structural properties.

\section{Analysis}

As noted in the Introduction, many galaxies are sufficiently unusual
that they cannot be classified in terms of the normal Hubble
scheme. This becomes increasingly true at intermediate redshifts. The
compact, luminous emission-line galaxies in paper II are such an
example. This is not due to poor spatial resolution, but to truly
unusual morphological properties, e.g., off-centered nuclei, tails,
asymmetric envelopes, etc.  To compare such objects morphologically to
``normal'' galaxies, we define here six fundamental parameters of
galaxy type that are quantitative, can be reliably determined over a
range in redshift, and are physically meaningful.

Two of these parameters are photometric, derived from existing
ground-based imaging and estimated $k$-corrections: rest-frame color
\bv, and absolute blue luminosity $M_B$. Two are image structure
parameters, derived from multi-aperture photometric analysis presented
below: physical half-light radius $R_e$, and image concentration
$C$. One is a combined photometric-structural parameter: average
rest-frame surface brightness $SB_e$ within $R_e$. Of the three
parameters luminosity, half-light radius, and surface brightness, any
one can obviously be derived from the other two. (We consider all
three since in any given range of, e.g., luminosity, there is
significant dispersion in both $SB_e$ and $R_e$.) The sixth parameter,
a 180$^{\circ}$-rotational asymmetry index ($A$), utilizes the
multi-aperture photometry indirectly through definition of the
extraction radius for rotation; we refer to $A$ as a structural
parameter.  Table 2 contains all individual measurements for
the Frei \etal sources. Luminosities and all image-structure
parameters are measured in the rest-frame $B$ band.

\subsection{Photometric parameters: restframe color and luminosity}

While the Frei \etal (1995) data set contains $B_J$ images for 75\% of
the sample and $g$ band images for the remaining objects, there is no
blue bandpass in which observations are available for all galaxies
(see Table 1). We have made a comparison of the apparent (uncorrected)
$B$ magnitudes listed in the {\it Third Reference Catalogue of Bright
Galaxies} (\cite{vau91}, RC3) to those derived from our photometry of
the Frei \etal images, appropriately transformed to the $B$-band using
the tabulated corrections of Frei and Gunn (1994). This comparison
shows that while the two magnitude estimates do not differ in the
mean, there is a 0.25 mag (rms) scatter. To avoid the uncertainty
associated with SED-dependent color transformations (see also \S A.3)
we use the RC3 uncorrected $B$ magnitudes and \bv \ colors instead of
the values from our own photometry.

We apply $k$-corrections and corrections for galactic extinction to
the \bv \ colors and apparent $B$ band magnitudes of the nearby galaxy
sample in the manner described in RC3. The heliocentric velocities
$v_{hc}$ of the galaxies are small (no greater than 3000 km s$^{-1}$
for any object); the average velocity is $\langle v_{hc} \rangle \sim
1000$ km s$^{-1}$. Hence the associated $k$-corrections are $<0.05$
mag. We use the distances given in the {\it Nearby Galaxies
Catalogue} (\cite{tul88}), recalculated to $H_0$=50 km s$^{-1}$
Mpc$^{-1}$, to derive absolute magnitudes $M_B$ from the corrected
apparent magnitudes. Note we do not correct for internal extinction
since the suitability and procedure for applying such corrections
may be ill-defined for higher-redshift galaxies.

\subsection{Multi-aperture photometry}

To characterize the light distributions of the galaxies, we performed
multi-aperture photometry on all images.  The apertures are centered
at the intensity-weighted centroid of each object. Since much of the
profile shape information is contained in the central parts of the
image, logarithmically spaced apertures are used. For the photometry
of the Hubble Space Telescope ($HST$) images in paper II, the smallest
aperture corresponds to 0.\arcsec05, the largest to 15\arcsec; for the
nearby galaxy photometry here, the aperture radii are scaled to
correspond to similar linear sizes. The apertures are circular to
accommodate the irregular morphology of the intermediate redshift
galaxies in paper II that would be difficult to fit with another
geometrical figure. The efficacy of this approach is addressed in more
detail below and in the Appendix (\S A.3).

\subsection{Structural parameters}

Image structure is most commonly quantified via bulge--disk
decomposition, yielding a bulge-to-total ratio, $B/T$. We refrain from
this approach here, for reasons which we alluded to in the
Introduction. For example, $B/T$ parameter may be poorly defined for
asymmetric and compact galaxies. Irregularities in the surface
brightness profiles, which can be caused by asymmetric structure,
rings, or lenses, also cause problems for bulge--disk decompositions.
While Kent showed that the concentration parameter correlates well
with the bulge-to-total ratio, this holds only for objects with $B/T <
0.63$. At larger values of $B/T$, bulge-disk decomposition fails for
several objects in Kent's sample, resulting in galaxies of type S0 --
Sa being given extremely high values of $B/T$. Bulge--disk
decomposition also becomes unreliable when galaxy disks are fainter
than the bulges.

It is worth noting again that these problems mainly arise from older,
one-dimensional methods of decomposition. The newer two-dimensional
decomposition techniques are clearly successful at reproducing the
observed light profiles, with remarkably small residuals (Schade \etal
1995, 1996; \cite{sim98}; \cite{mar98}). Still, there are physical
situations where bulge--disk decomposition techniques in general
become problematic, namely where the astrophysical reality is more
complex than simple bulge--disk models.  Some galaxies have central
condensations better described by an exponential profile rather than
an $r^{1/4}$-law (\cite{wyse97}); many galaxies have strong
bi-symmetries, such as bars; virtually all galaxies have varying
degrees of asymmetry due to star formation, dust, or large-scale
gravitational perturbations and lopsidedness. All of these features
represent details that decomposition into bulge and disk components do
not address correctly. Simple disk and bulge decomposition is also
inadequate for disk galaxies where the luminosity profile deviates
from a pure exponential (\cite{fre70}), e.g. type I and type II
disks. (Type I disk profiles have an added component which contributes
to the light just outside the bulge region; the surface brightness of
a type II profile shows the opposite behavior (an inner truncation),
and drops below the level of an exponential profile in the region near
the center.)

Given the astrophysical complexity of real galaxies, the physical
interpretation of the derived model parameters of disk-bulge fits
remains uncertain. Nonetheless, such profile-fitting methods should be
useful for estimating non-parametric structural and photometric
parameters (e.g. characteristic sizes, surface brightness, image
concentration, and ellipticity) in a way that uses the data in an
optimal manner. In the current effort, however, we have taken a
completely non-parametric approach of measuring sizes,
surface-brightness, image concentration and asymmetry using
multi-aperture photometry rather than deriving a model-dependent $B/T$
parameter.

\subsubsection{Half-light radii and surface-brightness}

We define first our working definition of a total magnitude since it
represents the critical zeropoint for measurement of the half-light
radius and surface-brightness. We use the dimensionless parameter
$\eta$ to define the total aperture of the galaxies -- a limiting
radius which is {\it not} based on isophotes.\footnote{Isophotal radii
introduce redshift-dependent biases unless careful consideration and
corrections are made for dimming due to the expansion ($\propto
(1+z)^{-3}$ in broad-band photon counts) and $k$-corrections. While
such redshift-dependent biases are not an issue for the samples
studied in this paper, in future papers in the series this would be an
issue were we not to avoid isophotes.} The concept of defining the size
of a galaxy based on the rate of change in the enclosed light as a
function of radius was first introduced by Petrosian (1976). In terms
of intensity, $\eta$ can be defined as the ratio of the average
surface brightness within radius $r$ to the local surface brightness
at $r$ (\cite{djo81}; \cite{san90}). Like Wirth \etal (1994), we
follow Kron's (1995) suggestion to use the inverted form, $\eta (r)
\equiv I(r)/\langle I(r)\rangle$, which equals one at the center of
the galaxy and approaches zero at large galactic radii. The radius
$r(\eta=0.5)$ corresponds roughly to the half-light radius $r_e$.

Since $\eta$ is defined as an intensity {\it ratio}, it is not
affected by the surface brightness dimming effect that makes the use
of isophotes problematic. Moreover, $\eta$ is only dependent on the
surface brightness within a given radius and not on any prior
knowledge of total luminosity or the shape of the light profile. These
properties make it advantageous for faint object photometry. We
defined the ``total'' aperture of the intermediate-redshift objects as
twice the radius $r(\eta=0.2)$. The apparent total magnitudes are then
defined within this aperture. For ideal Gaussian or exponential
profiles, the magnitude $m_{0.2}$ within the radius $2r(\eta=0.2)$ is
approximately equal to the true total magnitude $m_{tot}$; more than
99$\%$ of the light is included with the radius $r(\eta=0.2)$. For an
$r^{1/4}$-law profile, there is a difference $m_{0.2} - m_{tot} \sim
0.13$ mag; this is due to the slow decline in luminosity at large
radii that characterizes this profile.  The radius $r(\eta = 0.2)$ was
chosen based on visual inspection of the curves of growth, derived
from the aperture photometry, out to large radii.

For reference, the theoretical value for the ratio of $r(\eta=0.2)$ to
half-light radius is 2.16, 1.95, and 1.82 for three standard profiles:
Exponential, Gaussian, and $r^{1/4}$-law, respectively. The observed
ratio is 2.3 $\pm$ 0.3 for \bv $< 0.85$ (with little trend with
color), but rises slightly (2.6 $\pm$ 0.25) for the reddest galaxies
with \bv $> 0.85$. A contributing cause to this rise is that for
about half of the reddest objects, $r_{1/2}$ has been underestimated
by $\sim 20\%$ because of their higher ellipticity. As we show in the
Appendix (\S~A.3.2), the half-light radii of early-type galaxies with
axis ratio $a/b > 2$ are systematically underestimated by up to
30$\%$. This effect will also cause small changes to the measured
image concentration (\S~3.3.2) of these galaxies.

A weak downward trend can be seen from blue towards red colors; 
this is what we expect since bluer objects tend to have exponential 
luminosity profiles, and redder objects are better described by 
$r^{1/4}$-law profiles. However, this trend is broken by the reddest 
objects ($\bv > 0.87$), which have higher values of $r(\eta=0.2)/r_{1/2}$ 
than what is expected for an $r^{1/4}$-law profile.

Finally, the angular half-light radii $r_e$ were determined from the
normalized curves of growth. Based on $M_B$ and (corrected) $R_e$ we
calculated the photometric-structural parameter $SB_e$, the average
blue surface brightness within the half-light radius, for all objects.
For the nearby galaxy sample, the Tully catalog distances (as
described in \S~3.1) were used to determine $R_e$ (kpc).

\subsubsection{Image concentration} 

We use the image concentration parameter $C$ as defined by Kent
(1985), which is based on the curve of growth. This parameter was
shown to be closely correlated with Hubble type for ``normal''
galaxies: \begin {eqnarray*} C \equiv 5\,log(r_{o}/r_{i}) \end
{eqnarray*} In the above equation, $r_o$ and $r_i$ are the outer and
inner radii, enclosing some fraction of the total flux. In contrast,
the concentration parameter defined by Abraham \etal (1994) is not
based on curve of growth radii, but on a flux ratio within two
isophotal radii.

However, in practice Kent also uses isophotes: He replaces the outer
radius $r_o$, which encloses 80$\%$ of the total light, by the radius
of the 24th mag/arcsec$^2$ isophote. He has demonstrated that this
radius encloses $\sim 79\%$ of the total light for all galaxy types in
the restrictive confines of his sample (\cite{ken84}). Because of the
surface brightness dimming effect that becomes important for non-local
galaxies, we instead use a method that is independent of isophotes.
The total aperture of the galaxy, which determines the curve of
growth, is defined based on the $\eta$-radius as described in
\S~3.3.1.

We have also explored the possibility of using $\eta$-radii to define
a concentration parameter. However, a concentration parameter based on
the curve of growth was ultimately found to be the more robust
measure: the curve of growth increases monotonically with galactic
radius for all objects, while the $\eta(r)$-function will be
non-monotonic for a ``bumpy'' light profile (like that of a
well-resolved spiral galaxy). As a consequence, image concentration
defined by the curve of growth rather than $\eta$ exhibits less
scatter when plotted against other correlated observables (e.g. color,
surface-brightness) than an image concentration parameter based on the
$\eta$-function.

Anticipating our need to measure image concentration for small
galaxies in paper II and future papers in this series, we have studied
the effects of spatial resolution and $S/N$ on $C$.  Here we focus
primarily on resolution, as this was the dominant effect. The
importance of resolution is demonstrated by the comparison of Schade
\etal (1996) of decompositions of compact objects in ground-based and
$HST$ images: the cores of the blue nucleated galaxies are not
resolved in ground-based imaging, and hence they are frequently
misclassified as having much lower $B/T$-ratios than what is revealed
by $HST$-imaging. In paper II we analyze this sample of galaxies, and
hence this illustration is of particular relevance.

Resolution effects on image concentration were estimated by
block-averaging the images of nearby galaxy sample over a range of
values until the spatial sampling (as measured in pixels per
half-light radius) was comparable to that of the compact galaxies at
intermediate redshift observed with the WFPC2. The details of these
simulations are presented in the Appendix (\S A.1). In short, as the
objects' half-light radii get smaller, the scatter in the measured
concentration indices increases.  While larger inner radii or a
smaller outer radii decrease this scatter (due to improved resolution
and $S/N$, respectively), such choices decrease the dynamic range of
the concentration index.

Based on these simulations, we chose a definition of $C$ that is, to
first order, sufficiently robust to allow a direct comparison of the
image concentration of the local and the higher-redshift samples
studied here and in paper II, and furthermore gives a large dynamic
range: $C = 5\,log(r(80\%)/r(20\%))$. This concentration index is
remarkably stable: The mean concentration does not deviate from that
measured in the original image by more than 0.2, or $\sim 8\%$ of the
dynamic range in $C$, down to resolution of five pixels per half-light
radius.

Our definition is sufficiently close to that of Kent's (1985) so that
it is meaningful to compare our values directly to those he determined
from photometric analysis of a sample of nearby galaxies.  With this
choice of radii, a theoretical $r^{1/4}$-law profile has $C=5.2$, an
exponential profile has $C=2.7$, and a Gaussian has $C=2.1$.  These
values agree well with the results of Kent's analysis: he finds that
elliptical galaxies have $C \sim 5.2$, and late-type spirals have $C
\sim 3.3$.

Lastly, since we use circular apertures, the measured image
concentration may be affected by the ellipticity of the galaxy. Based
on the comparison between our results for the Frei \etal sample and
those of Kent's elliptical aperture photometry, we believe this to be
a negligible effect in all cases but the earliest, must elliptic
galaxies.  Wirth, Koo and Kron (1994) found that for an $r^{1/4}$ law
profile with axis ratio $b/a=0.2$, the change in $C$ is less than
5$\%$. The effect appears to be larger in our study. A more detailed
description of this possible systematic is given in the Appendix.

\subsubsection{Image asymmetry} 

The last image structure parameter is rotational asymmetry, $A$, as
defined by Conselice, Bershady \& Jangren (2000). This definition
differs from earlier methods in that the asymmetry is determined
within a constant $\eta$-radius of $\eta=0.2$, a noise correction is
applied, and an iterative procedure which minimizes $A$ is used to
define the center of rotation. This algorithm was tested to be robust
to changes in spatial resolution and signal-to-noise by Conselice
\etal (1999) using simulations similar to those described here for the
concentration parameter $C$; the systematics with resolution are below
10$\%$ of the original value for galaxies in paper I and II here.

\subsubsection{Morphological $k-$corrections}

To obviate the issue of `morphological' k-corrections, image
structural parameters should ideally be measured at the same
rest-frame wavelength for all objects. Anticipating our needs to
derive the structural parameters for intermediate-redshift objects in
paper II (and future papers in this series), we have adopted the
following protocol: (i) For the nearby galaxy sample we use the images
in the $B_J$ and $g$ bands to derive the primary local image structure
parameters. (The rest-frame wavelengths sampled by the $R,r$ band
images correspond to bands redshifted into the near-infrared for the
intermediate-redshift galaxies.) (ii) We use the multi-band
images of the Frei \etal sample to determine corrections to compensate
for the wavelength dependence of asymmetry, concentration, and
half-light radius -- as described in the Appendix (\S~A.2). For
example, $HST$ Wide Field Planetary Camera-2 (WFPC-2) images in the
$I_{814}$ band of objects between $0.3<z<0.8$ correspond to first
order to the rest-wavelength range of the $B_J$ and $g$
bands. Nonetheless, the effective rest-wavelength for such
intermediate-redshift galaxies is typically slightly redward of
rest-frame $B_J$ and $g$ bands. The corrections in \S~A.2 are suitable
for such samples, as well as higher redshift samples imaged in redder
bands.

\section{Results}

\subsection{Mean Properties, Distributions and Correlations}

The mean properties for our six parameters (M$_B$, $B-V$,
$R_e$, $SB_e$, $C$, and $A$) are listed in Table 3, as a function of
Hubble Type. While we would like to move away from using `Hubble
Types,' they are so ingrained in the astronomical culture that they
are a useful point of departure. For clarity in the following
discussion, we group these types together into ``Early'' (E-S0),
``Intermediate'' (Sa-Sb), and ``Late'' (Sc-Irr). These names are
potentially misleading, of course, and so we encourage the reader to
treat them as labels which evoke, at best, a well-conceived galaxy
type, but not necessarily an evolutionary state. Clearly further
sub-division could be made, but our current purposes are illustrative,
not definitive.

A typical approach to exploring the correlations in (and
dimensionality of) a multivariate parameter space is principal
component analysis. While this is valuable, it is not particularly
instructive for a first understanding of the distribution of different
types of objects in the parameter space. We are interested both in
correlations between observables and in trends as a function of the
qualitative Hubble-type. These correlations and trends need not be one
and the same. For example, two observables can be uncorrelated but
still exhibit a distribution segregated by Hubble type. To develop
such an understanding, we therefore inspected the 15 possible
2-dimensional projections of our 6-dimensional parameter space.


To distill this information further, we considered that there are in
fact three types of physically-distinct parameters:
\begin{enumerate}

\item spectral index (color): this parameter is purely photometric, by
which we mean there is no information about the shape of the light
profile. There is also no scale information, i.e. the amplitude and
size of the light profile is also unimportant. In the balance of this
paper we will use ``color'' and ``spectral index''
interchangeably.

\item form ($A$, $C$): these parameters are purely structural, by
which we mean that they do not depend -- to first order -- on the
amplitude or the shape of the spectral energy distribution, nor on the
physical scale of the light distribution; they reflect only the {\it
shape} of the light profile.\footnote{We consider image concentration
to be a form, in contrast to Morgan who used it as a surrogate for
spectral index.}

\item scale ($R_e$, $L$, and $SB_e$): these parameters are physically
distinct.  Luminosity is purely photometric (by our above
definition). Size, which we also refer to as a structural parameter,
is influenced by image shape, i.e., depending on the definition of
size, two galaxies with different light profile {\it shapes} can have
relatively different {\it sizes} (see \S3.3.1, for example).
Surface-brightness is a hybrid, photometric {\it and} structural
parameter; it is a function of size and luminosity. While
surface-brightness is a ratio of luminosity to surface area, it is still
a measure of ``scale'' -- in this case, the luminosity surface-density.
\footnote{A fourth scale parameter which we do not consider here is
line-width, or some measure of the amplitude of the internal
dynamics.}

\end{enumerate}
This reduces the types of combinations (by parameter-type) to 6,
i.e. between color, form, and scale.


We find the strongest and physically most interesting correlations are
between color, form, and the one scale parameter, $SB_e$ (Figures
3-5).  We focus on these for the remainder of the paper. Before
turning to them, for completeness we first summarize our observations
of the other types of correlations:

Color-color correlations are strong and well known (e.g. Figure
2). Effectively they add higher-order information about spectral
type. Here we consider only $B-V$ as a simple spectral index which
effectively represents the first-order information of spectral type.
In general, one might adopt several spectral indices, e.g. $U-V$ and
$V-K$, or a single index based on multi-colors.

Color-scale correlations also have been explored in detail elsewhere,
e.g.  color-luminosity relationships, known to exist for all galaxy
types in both the optical and near-infrared (Huchra, 1977b; Mobasher
\etal 1986; Bershady 1995). The limited dynamic range of the Frei
\etal sample in size and luminosity (they are mostly large and
luminous systems) preclude useful results being drawn here in this
regard. For example, the correlation of color with size in this sample
is subtle and depends in detail on how size is defined, as noted
above. Form-scale correlations including size and luminosity are also
difficult to assess for this sample for the same reasons of limited
dynamic range in scale.  However, scale versus scale {\it is} an
interesting diagnostic because, for example, size and luminosity allow
one to probe the range of surface-brightness in the sample. We explore
this in paper II.

\subsubsection{Spectral index versus form and scale}

Strong correlations exist in all three plots of color versus form
parameters $C$ and $A$ and scale parameter $SB_e$ (Figure
3). Early-type galaxies are redder, more concentrated,
high-surface-brightness, and more symmetric than Intermediate- and
Late-type systems.  The best correlation is between color and
concentration in the sense that there is a smooth change in both
quantities with Hubble Type. This is expected from a simple
interpretation of the Hubble Sequence as a sequence parameterized by
the relative dominance of a red, concentrated bulge (or spheroid)
versus a bluer, more diffuse disk.\footnote{A few of the local
galaxies have values of $C$ that are lower than the theoretical
concentration for an exponential disk (the errors in $C$ are $\lesssim
0.02$ for all of them). The majority of these objects are late-type
spiral galaxies with prominent, bright regions of star formation in
the spiral arms. The star-forming regions cause the image profiles to
become less centrally concentrated than a simple disk profile.} In
contrast, the distinction between Hubble Types in $SB_e$ and $A$ is
most pronounced between Early-types and the remainder; Intermediate-
and Late-types galaxies are not well distinguished by either of these
parameters.

A more complete local sample will likely include a larger
fraction of objects that do not follow these trends. For example,
amorphous galaxies have surface brightnesses comparable to elliptical
galaxies but are generally quite blue in color (\cite{gal87};
\cite{mar97}). Nonetheless, what is physically compelling about these
color-form correlations is that each axis carries distinct
information, respectively, on the integrated stellar population and
its spatial distribution. 

\subsubsection{Form versus form and scale}

There are clear trends present in the two plots of form versus $SB_e$
(scale) in Figure 4 as well as the plot of form parameters along in
Figure 5.  More centrally concentrated galaxies have higher average
surface-brightnesses and lower asymmetry; more symmetric objects have
higher surface-brightness. In general, the concentrated, high
surface-brightness galaxies are Early-type, while the Late-type
galaxies are less-concentrated, have lower surface-brightness, and are
more asymmetric. While there is substantial scatter in the form and
scale parameters for Early and Late types, these two extreme groups
still are well-separated in the above three plots. The
Intermediate-type galaxies, however, are {\it not} well separated from
these extremes, and tend to overlap substantially with the Late-type
galaxies, consistent with what is found in plots of color versus form
and scale: Intermediate- and late-type galaxies have comparable
degrees of asymmetry, and similar surface-brightness.

One should be cautious in concluding the relative merits of form-scale
and form-form and versus color-form and color-scale correlations based
on the relative separation of Hubble Types. Using Hubble Types may be
unfair if, for example, they were designed to correlate well with
color but not necessarily with the quantitative form and scale
parameters explored here. Since the form-form and form-scale
correlations themselves are comparable, and nearly as strong as for
color-form and color-scale, we are inclined to consider both as part
of a general classification scheme. Certainly the form and scale
parameters will each have different sensitivity to stellar evolution
than color and so are advantageous to consider in isolation.

\subsubsection{Comparisons to previous work}

The correlation between image concentration and mean
surface-brightness within the effective radius (Figure 4) has been
explored by several groups in the context of galaxy classification
(\cite{oka84}; \cite{wat85}; \cite{doi93}; \cite{abr94}). We focus
here, however, on Kent's (1985) $r$-band study since his definition of
image concentration and effective surface-brightness are the most
similar to our own. While similar, nonetheless the slope of the
correlation is steeper for our sample, albeit with much larger
scatter, as illustrated in the top panel of Figure 6. As the middle
and bottom panels reveal, the cause of the steeper slope in our sample
is due to a smaller dynamic range in image concentration. This is
likely due to the fact that we use circular apertures when performing
surface photometry, whereas Kent used elliptical apertures. We attempt
to quantify the systematics due to differences in aperture shape in \S
A.3. While the dynamic range in image concentration is reduced using
circular apertures for the Frei \etal sample, there does appear to be
a somewhat smaller scatter in $C$ as a function of $B-V$.

The nature of the large scatter in the top two panels of Figure 6 for
the Frei \etal sample is also discussed further in \S A.3. In short,
we believe much of this scatter is due to uncertainties in the $R$- and
$r$-band zeropoints of the Frei \etal sample. These uncertainties
adversely affect only the surface-brightness values in Figure
6. Robust estimators of the scatter about a mean regression (i.e.,
iterative, sigma-clipping of outlying points) eliminate the outlying
points, but still yield 50\% larger scatter in $R$-band $SB_e$ for
the Frei \etal sample as a function of either image concentration or
$B-V$. A plausible additional source contributing to this larger
scatter is that Kent's observed surface brightnesses were converted to
face-on values, while ours were not ``corrected'' in this way. We
conclude that if accurate and appropriate inclination corrections are
possible to apply to all galaxies in a given study, this would be
desirable. Since such corrections cannot be performed for the
intermediate-redshift objects in paper II (and in general, if such
corrections are not possible for a critical subset of the data), we
believe it is best not make such corrections at {\it any} redshift.

The asymmetry--concentration plane has also been explored for galaxy
classification purposes by, e.g., Abraham \etal (1994, 1996a) and
Brinchmann \etal (1998). Our methods of measuring these parameters
differ from theirs, and thus our quantitative results cannot be
directly compared. However, a qualitative comparison to the $A-C$ plot
of Brinchmann \etal shows that both methods yield very similar
results: the distribution of galaxies can be subdivided into sectors
where early-type, intermediate-type, and late-type objects dominate.
Brinchmann \etal also use the local sample from the Frei catalog to
define these bins, but note however that the points they plot
represent a sample of intermediate-redshift galaxies.  The $A-C$
correlation in the Brinchmann \etal diagram is not as clear as that
seen here for the local sample in Figure 5; the scatter in
their diagram is comparable to the dynamic range of the parameters.
This is probably due to the different properties of the samples,
rather than to the differences in how we determine the parameters. For
a more direct comparison, we plot $B$ band asymmetry and concentration
versus rest-frame \bv \ color for 70 galaxies from the Frei \etal
sample (Figure 7), using both the $A$, $C$ values from this
study and those found by Brinchmann \etal \ It can be seen that the
distributions are overall quite similar; however, the separation in
asymmetry of the different Hubble types is more apparent in this
study, and the scatter in concentration is somewhat smaller.  The
conclusion here, then, is that our methodology offers typically
modest, but sometimes significant improvements over previous work.

\subsection{Classification}

The above results point to how we can most effectively define a
parameter demarcation to isolate, identify, and classify normal
galaxies. In the four-dimensional parameter space of ($B-V$, $A$,
$SB_e$, $C$), we define boundaries (``cuts'') in the 6 two-dimensional
projections between galaxies classified in the Hubble Sequence as
Early/Intermediate and Intermediate/Late. These boundaries, selected
by eye on the basis of the distribution of Hubble types, are listed in
Table 4 and illustrated in Figures 3-5. Segregation by
higher-dimensional hyper-surfaces are likely to be more effective
(galaxies appear to be distributed on a `fundamental' hyper-surface --
the subject of a future paper), but the projected boundaries here are
meant as illustrative, and practical for application when all of the
parameters are not available. We stress that these boundaries
are not definitive in some deeper physical sense.  For example, in
terms of formal Hubble Types cuts involving color are clearly ``best;''
however, as noted above, this may not be physically significant.

It would be uninteresting if all of the cuts provided the same
classification. Moreover, one expects there will be discrepancies for
objects near boundaries. We find that 49\% of the sample matches in
all cuts, while 64\%, 87\%, and 99\% of the sample matches in at least
5, 4, or 3 cuts, respectively. (Hereafter, we refer to cases where 5
out of 6 cuts match as ``5/6,'' etc.) This degree of consistency seems
reasonable so we have not tried to fine-tune the boundaries (such
fine-tuning would not be sensible anyway since the details of the
classification self-consistency are likely to be sample-dependent):
The preponderance of objects are classifiable by a simple majority of
the classifications based on the 6 cuts; 13\% of the objects have a
more ambiguous classification.

Of interest are the discrepancies within and between cuts in different
combinations of color, form, and scale. We found that it is useful to
group the six cuts into two groups of three. The first consists of the
cuts in Figure 3 between color, form and scale, which we refer to as
color-form/scale. The second consists of the cuts in Figures 4 and 5
between form and scale, which we refer to as scale/form-form.  For
example, 64\% of the variance in the 5/6 cases comes from cuts in
$C$--$SB_e$, whereas cuts in $C$--$(B-V)$ and $A$--$C$ are always
consistent with the majority classification. More generally,
scale/form-form cuts are internally mis-matched 40\% of the time,
while color-form/scale cuts are internally mis-matched only 21\% of
the time (and two-thirds of these color-form/scale mis-matches are
also present in scale/form-form mis-matches). In other words, the
color-form/scale cuts tend to be more consistent; much of the variance
in the scale/form-form cuts again comes from $C$--$SB_e$.

Only two galaxies pose a substantial problem for classification: NGC
4013 and NGC 4216. They are classified by various cuts to be in all
categories (Early, Intermediate, and Late), and have no majority
classification. However, both are highly inclined (4013 is edge on),
which appears to give them unusual observed properties. Indeed, they
are extreme outliers in several of the projections in Figures 3 and 4
(see also A.3.2 and figures therein). Hence such problem cases are
likely to be easy to identify. Three other sources classified in all
three categories (NGC 4414, 4651, and 5033) are not a problem: They
have 4/6 consistent classifications. Two of these (NGC 4651 and 5033)
have Seyfert nuclei, and are outliers only in plots with image
concentration; they are highly concentrated for their color. NGC 4414
is not an outlier in any of the plots.

Finally, it is interesting to note that 23\% of the sample has
inconsistent majority classifications in color-form/scale versus
scale/form-form cuts. This is true for 100\% of the 3/6 cases, and
55\% of the 4/6 cases. However, we believe this is for different
reasons. In the latter cases (only) we find that the galaxies are
predominantly at high inclination ($\sim$50\% excess in the top half
and top quartile of the sample distribution in inclination). Moreover,
the color-form/scale classifications in these cases are all {\it
earlier} than the majority scale/form-form classifications. We surmise
this is due to the effects of reddening on $B-V$.\footnote{Inclination
will also cause changes in other measured parameters. Changes in $C$,
however, appear to be small (see \S A.3.2). Surface-brightness will
tend to increase at modest inclinations, and then decrease at high
inclinations if a prominent dust lane obscures the bulge.  Likewise,
$A$ may increase due to a dust lane until the galaxy is directly
edge-on. As a consequence of these changes and the distributions and
cuts, $C$--$SB_e$ tends to mimic the color-form cuts in the
high-inclination cases, while $A$--$SB_e$ and $A$--$C$ do not.} While
the color-form/scale classifications tend to be earlier for the 3/6
cases, because there is no apparent inclination dependence, these
differences are due likely to other physical effects. Two
possibilities include low star-formation rates or high metallicity for
galaxies of their form. Both of these conjectures are testable via
spectroscopic observation.

We suggest then, as a practical, {\it simple} prescription, that the
majority classification for all 6 cuts be taken as the classifier,
except in the situation where the galaxy in question is highly
inclined. In the latter case, the majority classification of the
scale/form-form cuts should be adopted.  When galaxies have only 3/3
consistent classifications, (13\% of the Frei \etal sample), the
adopted classifier should be intermediate between the two most common
classifications. It also may be of interest to note if the
color-form/scale and scale/form-form majority classifications
differ. However, further elaboration based on these two-dimensional
projections of a higher-dimensional distribution is not likely to be
warranted.

\subsubsection{Discussion}

We note that there are no distance-dependent scale parameters in our
classification. By this we mean specifically that the classification
parameters do not depend on knowledge of the distance modulus. Hence
this classification is both quantitative and independent of the
cosmological distance-scale and its change with cosmological epoch
(i.e. no {\it a priori} knowledge is needed about H$_0$ or q$_0$).
The effects of the expansion do change the {\it observed}
classification parameters. However, with knowledge of galaxy redshifts
and judicious choice of ``redshifted'' photometric bands,
surface-brightness dimming can be corrected and band-shifting either
eliminated or corrected via the protocol described in the
Appendix. Galaxy evolution, of course, will also modify the values of
the parameters, but this is precisely the utility of the
classification systems as applied to such a study: In what way do the
parameters and their correlations evolve?  How do the scale parameters
change for a fixed range in classification parameters?  These are
issues which we intend to explore in subsequent papers in this series.

We also comment on the efficacy of using the four-dimensional
parameter space of color, concentration, surface-brightness, and
asymmetry for the classification of distant galaxies.  As noted
earlier, Abraham \etal (1996a) and Brinchmann \etal (1998) have
explored the use of the asymmetry--concentration plane as a tool for
distant galaxy classification. The use of the additional parameters of
color and surface-brightness are clearly advantageous; they offer
substantially more information, particularly as a diagnostic of the
stellar population age and surface-density. The reasoning behind using
$A$ and $C$ alone has been that to first order, they can be estimated
without redshift information. Yet the wavelength dependence of both
parameters (i.e., what is referred to as `morphological
k-corrections') can lead to measurement systematics. These
systematics, if not corrected, in turn result in objects over a range
in redshift to be systematically misclassified. For example,
Brinchmann estimates that at $z = 0.9$, $25\%$ of spiral galaxies are
{\it mis}-classified as peculiar objects in the $A - C$ plane. This
fraction is expected to increase at larger redshifts. Hence, for
high-$z$ studies of galaxy morphology, redshift information is crucial
even when using asymmetry and concentration. Therefore, since
redshift information is crucial no matter what, there is no reason
{\it not} to use the four-dimensional classification we have outlined
in future studies. The recent refinements and calibration of the
technique of estimating redshifts photometrically make this all the
more tractable.

Finally, we note that while the classification we have proposed here
is practical and useful, there are five areas where we anticipate it
can be improved or elaborated: (i) As we have mentioned before, the
spectral-index parameter could have much greater leverage at
distinguishing between different stellar populations by adding
pass-bands that expand the wavelength baseline (e.g. the $U$ and $K$
bands in the near-UV and near-IR, respectively), or by increasing the
spectral resolution (e.g. line-strengths and ratios). A further step
of elaboration would be to explore spatially resolved spectral indices
(gradients) and determine their correlation with form parameters.
(ii) Internal kinematics should be considered. Ideally, the kinematic
information would include estimates of both the random and ordered
motion (rotation) so that the dynamical temperature could be assessed,
in addition to the overall scale. Kinematics are relatively expensive
to obtain (compared to images), but with modern spectrographs on large
telescopes, the absolute cost is minimal at least for nearby
galaxies. (iii) Higher-dimensional correlations are worthy of
exploration to determine, for example, whether ``fundamental''
hyper-planes can adequately describe the entirety of the galaxy
distribution. (iv) It is worth considering whether there are
additional form parameters of value for classification that have not
been included here. (v) The classification scheme needs to be tested
against much larger, and more volume-representative samples of
galaxies.

\section{Summary and Conclusions}

We have presented results from a study of the photometric and
image-structural characteristics and correlations of a sample of
local, bright galaxies (Frei \etal 1996). We find it illuminating to
distinguish between parameters which characterize spectral-index
(color), form (image concentration and asymmetry), and scale (size,
luminosity, and surface-brightness). In this context, we arrive at the
following main results and conclusions.

\begin{itemize}

\item We find that a combination of spectral-index, form and scale
parameters has the greatest discriminatory power in separating normal
Hubble-types. The strongest correlation is found between color and
image concentration. However, there are equally strong correlations
between form parameters (e.g. $A$ and $C$), but here the Hubble-types
are not as well distinguished. As an indicator of classification
utility, we suggest that the strength of the correlation between
parameters is likely more important than the separation of Hubble
Types within the correlation.

\item It is possible to define a quantitative classification system
for normal galaxies based on a four-parameter sub-set of
spectral-index, form and scale: rest-frame \bv \ color, image
concentration, asymmetry, and average surface brightness within the
half-light radius.  We propose a specific classification that
distinguishes between ``normal'' galaxies as Early, Intermediate, and
Late based on cuts in these four parameters.  The classification is
successful for 99\% of the Frei \etal sample.  Nonetheless, we
designate this as ``preliminary'' until larger, more comprehensive
samples of galaxies are needed than analyzed in the present study.

\item Distance-dependent scale parameters are {\it not} part of this
preliminary classification.

\item These classification parameters can be measured reliably over a
broad range in $S/N$ and image resolution, and hence should be
applicable to reliably distinguishing between a wide variety of
galaxies over a large range in redshift.

\item Redshift information {\it is} needed to estimate reliably both
the photometric properties (rest-frame color and surface brightness)
as well as the structural parameters asymmetry and concentration at a
fixed ($B$-band) rest-frame wavelength. In terms of redshift
independence, asymmetry and concentration alone thus offer no
advantages over the additional parameters classifiers proposed
here. Indeed, incorporating the full suite of parameters defined here
is advantageous for the purposes of classification.

\end{itemize}

\acknowledgements

The authors wish to thank Greg Wirth for his highly-refined algorithm
for calculating $\eta$-radii user here and in paper II; our
collaborators David Koo and Rafael Guzm\'an for their comments on the
manuscript and input on assembling dwarf galaxy samples; and Jarle
Brinchmann for providing us with his measurements of structural
parameters for the local galaxy sample. We also gratefully acknowledge
Jay Gallagher and Jane Charlton for a critical reading of the original
version of this paper (I and II), and for useful discussions on this
work. Most importantly, we thank Zolt Frei, Puragra Guhathakurta,
James Gunn, and Anthony Tyson for making their fine set of digital
images publicly available. Funding for this work was provided by NASA
grants AR-07519.01 and GO-7875 from STScI, which is operated by AURA
under contract NAS5-26555. Additional support came from NASA/LTSA
grant NAG5-6032.

\begin{appendix}

\section{A. Corrections for Measurement Systematics}

Here we establish the measurement systematics due to changes in image
resolution for half-light radius and image concentration, and for
band-shifting effects on half-light radius and image concentration,
and asymmetry.  Systematic effects of images resolution and noise on
asymmetry are quantified in Conselice \etal (1999).

\subsection{A.1. Resolution dependence of observed size and image concentration}

To maximize the dynamic range of the measured concentration index,
$C$, the inner radius should be small, and the outer radius large
relative to the half-light radius. In this way, one samples the light
profile gradients in both the central and outer regions of a galaxy
where the bulge and disk contribute quite differently.  This strategy
maximizes the leverage for discriminating between different profiles,
e.g. exponential and $r^{1/4}$-law. In the presence of noise and
limited spatial resolution, however, the choice of radii determines
the robustness of the concentration index: As noted by Kent (1985),
the inner radius should be large enough to be relatively insensitive
to seeing effects, and the outer radius should not be so large that it
is affected by uncertainties in the sky background and $S/N$. In the
current study, the sources are resolved, and the images are at
moderately high signal-to-noise: within the half-light radius, the
sample of local galaxies have $600 \lesssim S/N \lesssim 3000$. The
intermediate redshift galaxies in paper II have $S/N$ in the range 40
to 90, with a mean of $\sim 55$. This is sufficiently high that we
focus our attention here on the effects of spatial sampling and
resolution.

Even in the absence of significant image aberrations, an additional
limiting factor is the number of resolution elements sampling the
inner radius. This is likely to become a limiting factor when the
half-light radii is only sampled by a few pixels. To understand this
potential systematic, we have calculated six concentration indices $C
= 5\,log(r_{o}/r_{i})$ for several different choices of inner and
outer radii.  We use $r_i$ enclosing 20 and 30$\%$ of the light, and
$r_o$ enclosing 50, 70, an 80$\%$ of the light. The radii were
measured for nearby galaxies that were block-averaged by factors 2, 4,
and 6 to simulate coarser spatial sampling, as shown in Figure
8. The six different concentration indices are plotted as a
function of sampling in Figure 9. These simulations span
sufficient dynamic range in size to cover most galaxies observed, for
example, in the Hubble Deep Field. With factors of 4 and 6, we
measure radii with pixel sampling similar to that observed in the
HST/WFPC-2 images of the intermediate-redshift objects of paper
II. Typically, these galaxies have half-light radii of 0.3 --
0.7\arcsec. For the Planetary Camera, the scale is 0.046\arcsec/pixel,
and for the Wide Field, 0.10\arcsec/pixel; hence the half-light radii
are of order 3 to 15 pixels.

The half-light radius $r(50\%)$ is remarkably stable, even with poor
sampling. Unfortunately, the dynamic range given by concentration
indices with $r_o = r(50\%)$ is too small to be useful. As expected,
the 30$\%$ radius was more stable than the 20$\%$ radius to 
decreased spatial resolution. However, the concentration indices using
$r(30\%)$ were less sensitive to the differences between galaxy types,
and gave a smaller dynamic range than indices using $r(20\%)$. The
inner radius dominated the effect on the amplitude of the systematics;
changing the outer radius from $70\%$ to 80$\%$ decreased the scatter
only marginally. With a block-averaging factor of 6, where the
half-light radii are typically only $\sim 5$ pixels, the scatter
becomes large for all choices of concentration indices.

Based on these simulations, we decided to use the radii enclosing
80$\%$ and 20$\%$ of the total light (as did Kent, 1985), even though
$r_o = r(70\%)$ gives concentration indices with slightly smaller
scatter at poor resolution. For objects with half-light radii of only
7 pixels, the mean differences in concentration (relative to the
original image) are $\Delta C_{80:20} = -0.10_{-0.60}^{+0.20}$ and
$\Delta C_{70:20} = -0.10_{-0.50}^{+0.15}$. Even at a resolution of
only five pixels per half-light radius, the concentration index only
deviates by 0.2 relative to the original image; this is $\sim 8\%$ of
the dynamic range in $C_{80:20}$. Thus we consider this parameter to
be robust enough to useful in the comparison of local and
intermediate-redshift samples.

\subsection{A.2. Systematics with wavelength} 

Observations at different wavelengths sample preferentially different
stellar populations in a galaxy. Since these populations are not
always spatially homogeneous, the image-structural characteristics
(concentration, asymmetry, and half-light radius) will will have some
wavelength dependence [e.g., see de Jong's (1995) study of disk
scale-lengths]. Hence, when comparing one of these parameters for
different galaxies, the parameter ideally should be measured at the
same rest-frame wavelength for all objects. This is not possible in
general for studies over a wide range in redshifts employing a finite
number of observed bands.  To determine the amplitude of the
wavelength-dependence for the measured structural parameters, we
therefore compare the $B_J$ and $R$ structural parameters for 72 of
the Frei \etal galaxies.  The differences between the red and blue
structural parameters versus the rest-frame color, \bv, are shown in
Figure 10. For comparison, all intermediate-redshift objects in
paper II, except two, fall in the bluest bin ($\bv < 0.62$).

The plot of $\Delta C = C_B - C_R$ shows that in most cases, the
values are slightly negative, i.e. the majority of objects are more
highly concentrated in the red band than in the blue, as expected
because of the redness of the central bulge. Only the bluest galaxies
have comparable image concentration in both bands. There is a weak
trend towards more negative values for the redder (early-type)
objects, which also show a larger scatter than the bluer objects.

In the plot of $\Delta A = A_B - A_R$ it is clear that most galaxies
have at positive values, i.e. their image structures are more
asymmetric in the blue band than in the red, as shown by Conselice
(1997). The difference in asymmetry is only seen for late- and
intermediate-type objects; red objects are generally very symmetric in
both bands, and have $A_B-A_R \sim 0$. This trend was also noted by
Brinchmann \etal (1998).

The plot of half-light radii ($\Delta R_e = R_{e,B} - R_{e,R}$) shows
that most values are slightly positive, with a larger scatter for
redder objects. No other trend with color is seen. The fact that the
galaxies have slightly larger half-light radii in the blue band is
consistent with their image concentration being higher in the red
band, as a bulge profile generally has a much smaller scale length
than an exponential profile.

In summary, the average differences ($\pm1\sigma$) between parameters
for galaxies with $\bv<0.62$, determined from the $B_J$ and $R$ bands,
are: $\Delta C=-0.15 \pm 0.30$, $\Delta A = 0.013 \pm 0.044$, and
$\Delta R_e=0.21 \pm 0.80$ kpc. 

\subsubsection{A.2.1. Corrections for wavelength systematics}

Based on the mean values above, we correct the measured structural
parameters for galaxies at non-zero redshift to the rest-frame $B$
band values as follows, where for clarity we use the intermediate
redshift galaxies in paper II as an example. The structural parameters
of these intermediate-redshift objects generally were measured at
rest-frame wavelengths between $B_J$ and $R$, i.e., in the observed
WFPC2 $I_{814}$ band for $z\sim0.6$.  For ``normal'' galaxies, this
would cause us to overestimate $C$, and underestimate $A$ and
$R_e$. Hence we use the differences listed above and the redshift of
the objects to linearly interpolate the correction to the measured
values.  Specifically, for a given parameter and color bin, we use the
mean difference between values measured in the $B_J$ and $R$ bands,
and the position of the rest-frame wavelength relative to the $B_J$
band, to make corrections to the measured parameters. (Note that the
correction made to $R_e$ also affects the value of $SB_e$ in general.)
For some objects, the combination of observed band-pass and redshift
corresponds to rest-frame wavelengths slightly blueward of $B_J$. When
computing the corrections for these objects, we assumed that the
wavelength trends continue outside the $B_J$ -- $R$ wavelength
range. Overall, these corrections are small for objects in paper II,
while for higher-redshift objects we expect band-shifting effects to
become increasingly important.

We add a final, cautionary note that it is not certain the corrections
for intermediate-redshift objects should be made based on the
correlations we see for the nearby sample. When comparing the
observations in the bluer bands ($B_{450}$ or $V_{606}$) to those in,
e.g. the $I_{814}$ used in paper II, we find that most objects are
more concentrated in the blue band, and slightly larger in the red
band -- this is the opposite of what we see for the Frei \etal
sample.\footnote{Indeed, Huchra noted that the Markarian galaxies get
bluer toward their centers, reminiscent of the blue ``bulges'' seen in
the blue nucleated galaxies of paper II, yet in contrast to the color
gradients found for ``normal'' galaxies. This type of color-aperture
relation was also noted by de~Vaucouleurs (1960, 1961) for the latest
Hubble-type galaxies (Sm, Im).} For asymmetry, the trend is the same
for both samples (higher $A$ in bluer bands). The trends are not
directly comparable, however, to what we see in the local sample, as
the observations in the bluer bands correspond to rest-frame
wavelengths in the UV region for most intermediate-redshift
galaxies. For this reason, and since the small sample of
intermediate-$z$ objects poorly defines the variation in image
structure with wavelength, we adopt the more well-determined trends
seen for the Frei \etal sample to calculate the band-shifting
corrections.  These corrections based on local galaxy trends tend to
make the intermediate-redshift objects somewhat less ``extreme'';
their half-light radii become larger, their surface brightnesses
fainter, and their image concentrations lower. If instead we had based
our corrections on the trends seen within the intermediate-$z$ sample
of paper II, then this sample would be even more extreme relative to
the local galaxy sample. The corrections would then tend to shift the
positions of the intermediate-$z$ objects even farther from the nearby
galaxies in diagrams that include any of the parameters $R_e$, $SB_e$,
and $C$.

\subsection{A.3. Systematics with aperture shape}

\subsubsection{A.3.1. Comparison to elliptical aperture photometry}

Circular-aperture surface-photometry will yield systematic differences
in the measured structural parameters when compared to those derived
from elliptical-aperture surface-photometry. To assess this, we
compared our results for the Frei \etal catalog in $R$ and $r$ bands
to the results of Kent (1985) for a sample of local, Hubble-type
galaxies (Figure 6). Kent used elliptical apertures tailored to fit the
axis ratio and position angle of each isophote in galaxy images to
determine $r$ band image concentration and average surface brightness
within the half-light radius. 

As we detail in the figure caption, we have attempted to transform all
of the surface-brightness values to the Cousins $R$ band ($R_c$). For
each of the relations in Figure 6 we have characterized the slopes and
scatter about a mean regression using a simple linear, least-squares
algorithm with an iterative, sigma-clipping routine to remove outlying
points. Given the nature of the data, such an algorithm is not
statistically correct (see, e.g., Akritas and Bershady,
1996). However, given the potential photometric uncertainties
(discussed below) and the need for robust estimation, it is not
possible to formally implement more appropriate
algorithms. Nonetheless, the relative characterization of the slopes
and scatter between Frei \etal and Kent samples is useful.

As discussed in \S 4.1.3, the slope of the correlation between average
surface-brightness and image concentration is steeper for our study
than for Kent's because of a decreased range in image concentration in
our study. The effect (bottom panel of Figure 6) is such that the
bluest galaxies have comparable image concentration values in both
studies while the image concentration of the reddest galaxies differ
by as much as 1 unit in the mean (Kent's values are larger).  We
interpret this as likely to be the effect of different aperture
shapes. The results of our study of systematics with axis ratio
(below) support this conclusion. Surprisingly, there is no indication
that elliptical apertures give significantly different results than
circular apertures for intermediate- and late-type (disk dominated)
galaxies.

The larger scatter in the Frei \etal (1996) sample in the top two
panels of Figure 6 might lead one to conclude that the elliptical
apertures provide a superior measurement of effective
surface-brightness. However, much of the scatter is due to the subset
of the Frei \etal sample observed at Palomar Observatory. We believe
that zeropoint problems are the cause of much of this scatter,
consistent with discussion in Frei \etal concerning the difficulty of
photometric calibration. The bulk of the objects observed at Lowell
Observatory are consistent with independent $R_c$-band photometry from
Buta and Williams (1996), although there are some points that are very
discrepant. In general, the overlap is excellent in $SB_e$ and $B-V$
between the Kent sample, the Lowell subset of the Frei \etal sample,
and the subset of the Frei \etal sample with Buta and Williams'
photometry.

\subsubsection{A.3.2. Systematics with axis ratio}

A second approach to determine the systematic effects of aperture
shape on measured structural parameters was also used: we quantify the
degree to which ``normal'' galaxies with the same intrinsic morphology
but with different axial ratios $a/b$ will have different $C$ when
measured with circular apertures. The galaxies in the Frei \etal
catalog were divided into early-, intermediate- and late-type objects
(using the same bins as elsewhere in the paper), and we plot
concentration and half-light radius versus the logarithm of the axis
ratio (taken from the RC3 catalog).

In the image concentration plots (Figure 11), a weak trend can be seen
for the late-type galaxies (top panel), with slightly higher values of
$C$ for the more inclined objects. This effect, if caused by the shape
of the apertures, will lead us to overestimate the image concentration
by at most $\sim 0.1$ ($3\%$) for the nearly edge-on galaxies. We do
not expect this to be a problem for our analysis.  The two labeled
objects have unusually high values of $C$ for their morphological
type. One of them, NGC 5033, is known to be a Seyfert 1 galaxy; the
other, NGC 4651, is a suspected ``dwarf-Seyfert'' galaxy (Ho \etal,
1997).  For intermediate-type objects (middle panel), no trend is
observed.  The lowest $C$-value, which belongs to NGC 4013, could be
caused by the prominent dust lane in this object: the central light
distribution is divided into two parts, making it difficult to
determine the position of the center. Effects like these will likely
be more problematic for objects with high values of $a/b$. The highest
$C$-value in this plot is that of NGC 4216, which also is highly
inclined and has spiral arm dust lanes superimposed on the bulge. In
the bottom panel, a trend is observed for the early-type galaxies: the
concentration is lower for objects with higher $a/b-$ratio. This
effect will cause us to underestimate the image concentration of these
objects by $\sim 0.5$, or 10--15$\%$. This result agrees well with
what was seen in the comparison of the Frei \etal sample to Kent's
image concentration measurements, as described above. This leads us to
conclude that our circular aperture photometry will underestimate the
image concentration somewhat for elliptical/S0 galaxies. Again, there
is no indication that the aperture shapes lead to different results
for intermediate- and late-type galaxies.

In the plots of half-light radius $R_e$ versus $a/b$ (Figure 12), no
trends are seen for the intermediate- and late-type objects. For the
early-type objects, however, the measured half-light radii become
progressively smaller for increasing values of $a/b$. The trend is
weak; it will cause us to underestimate the half-light radii by at
most 30 $\%$ for objects with $a/b \sim 4$.  If this effect is real,
the derived surface brightness will be too bright by $\lesssim 0.7$
magnitudes for the most highly elliptical early-type galaxies.

\end{appendix}

\input bershady_tables.tex

\centerline{\bf Figure captions}

\bigskip

\noindent Fig. 1.--- Rest-frame \bv \ versus $M_B$ for the nearby
galaxy sample of Frei \etal (E-S0, Sa-Sb, and Sc-Irr) and Huchra's
(1977a) sample of normal Markarian galaxies (pluses). The dotted
outline indicates the approximate locus of dE/dSph galaxies. The
intermediate redshift samples from paper II are also plotted for
comparison: blue nucleated galaxies (BNGs) compact, narrow
emssion-line galaxies (CNELGs), and small, blue galaxies (SBGs). (The
two SBGs and the BNG that we ultimately determine not to be ``Luminous
Blue Compact Galaxies'' in paper II are shown as hatched symbols.)
Only a few Markarian galaxies and late-type galaxies from the Frei
\etal catalog share the extreme color--magnitude properties of the
intermediate-redshift objects. In this plot, and in Figures 2-6,
the vigorously star-forming galaxy NGC 4449 is labeled.
Characteristic random errors are indicated separately for the Frei
\etal sample and the intermediate-$z$ objects.

\bigskip

\noindent Fig. 2.--- Rest-frame \ub \ versus \bv \ for the sample
samples as in Figure 1.  The intermediate-redshift samples of paper II
largely overlap with the bluest Markarian galaxies, which extend
blueward the color-color relation seen for the ``normal'' galaxies
from Kent.

\bigskip
 
\noindent Fig. 3.--- Rest-frame $B$-band form and scale parameters
versus spectral index for the Frei \etal sample.  Top panel: Average
surface-brightness within the half-light radius ($SB_e$) versus
rest-frame \bv. Middle panel: Image concentration ($C$) versus
\bv. Bottom panel: 180-degree rotational image asymmetry ($A$) versus
\bv. Characteristic errors are given in the top-left corner of each
panel. Outlying objects are labeled and discussed in the text. Dashed
lines demark Early, Intermediate, and Late types in our classification
scheme. Symbols are by Hubble type, as defined in the key. Different
Hubble types are well distinguished, particularly in
color. Morphological types are also well separated in $C$, but only
the earliest types are well separated in $SB_e$ and $A$.

\bigskip

\noindent Fig. 4.--- Rest-frame $B$-band parameters of form versus
scale for the Frei \etal sample. Top panel: Image asymmetry ($A$)
versus average surface-brightness ($SB_e$). Bottom panel: image
concentration ($C$) versus $SB_e$. Outlying objects are labeled and
discussed in the text. Dashed lines demark Early, Intermediate, and
Late types in our classification scheme. The separation of
morphological types is less clear than in Figure 3, but the different
Hubble types are reasonably segregated.

\bigskip

\noindent Fig. 5.--- Form versus form parameters for the Frei \etal
sample: Rest-frame $B$-band image asymmetry ($A$) versus image
concentration ($C$). Outlying objects are labeled and discussed in the
text. Dashed lines demark Early, Intermediate, and Late types in our
classification scheme. The separation of morphological types is less
clear than in Figure 3, but is comparable to figure 5 where the
different Hubble types are reasonably segregated.

\bigskip

\noindent Fig. 6.--- Comparison of form, scale, and spectral index
correlations between Frei \etal and Kent samples. {\it Top panel:}
average $R$ band (Kron-Cousins) surface brightness within the
half-light radius, $SB_e(R_c)$, versus $R$- or $r$-band image
concentration, $C(R)$. {\it Middle panel:} $SB_e(R_c)$ versus
rest-frame \bv. {\it Bottom panel:} $C(R)$ versus rest-frame \bv. {\it
Structural parameters:} We measured half-light radius and image
concentration for the Frei \etal sample using their $R$ or $r$-band
CCD images and circular photometry apertures. Kent measured these
structural parameters using elliptical apertures on $F$-band CCD
images. {\it Photometric parameters:} The Frei \etal sample is
subdivided between objects observed at (a) Lowell Observatory (filled
squares), (b) Palomar Observatory (dotted-circles), and (c) an
overlapping subset of the Frei \etal sample with existing $R_c$-band
photometry from Buta and Williams (1996; outlined-triangles). For (a)
and (b) we used the zeropoints from the Frei \etal image headers
(DNATO\_BV), and transformations from Thuan-Gunn $r$ and Gullixson
\etal $R$ to Cousins $R_c$ from Frei \& Gunn (1994). We have
transformed Kent's photometry reported in the Thuan-Gunn $r$-band to
$R_c$ again based on transformations in Frei \& Gunn (1994); Kent
corrected surface brightnesses to ``face-on'' values. {\it
Regressions:} Lines indicate $\pm1\sigma$ about linear least-squares
fits to the correlations (dotted, Kent; dashed, Frei \etal) using an
iterative clipping method ($\pm2.5\sigma$ clip; 10 iterations). In the
top and middle panels only the Lowell subset of the Frei \etal sample
was used in the regressions. The substantial scatter in the Frei \etal
$SB_e(R_c)$ values we infer is due primarily to zeropoint
difficulties; we detect no noticeable systematics effects with
inclination in $SB_e(R_c)$. The difference in the correlation between
$SB_e(R_c)$ and $C(R)$ is largely due to the shallower trend in $C(R)$
with \bv \ for the Frei \etal sample. This may be due to differences
between circular versus elliptical apertures. While elliptical
aperture photometry provides greater dynamic range in $C(R)$, the
correlation of $C(R)$ with \bv \ has larger scatter.

\bigskip

\noindent Fig. 7.--- Form parameters and spectral index for 70
galaxies from the Frei \etal sample as determined by Brinchmann
\etal Top panel: $B$ band image concentration versus rest-frame \bv.
Middle panel: $B$ band asymmetry versus rest-frame \bv.  Bottom panel:
$B$ band asymmetry versus concentration. The asymmetry parameter was
determined in a very similar manner as our own and thus should have
comparable dynamic range. Since our $C$ parameter is logarithmic, we
plot the logarithm of the Brinchmann \etal $C$ values. These
plots are displayed so that they may be directly comparable to Figures
3 and 5. The trend in asymmetry for the different Hubble types is more
apparent in Figure 3 and 5. In the concentration--color plane, the
distributions are similar for both studies, although we find a smaller
scatter among the late-type galaxies, and a larger scatter among the
early-type objects.

\bigskip

\noindent Fig. 8.--- A representative subset of galaxy images from the
Frei \etal catalog, block-averaged by factors 1, 2, 4, and 6 (top to
bottom).  While the apparent change in qualitative (visually-assessed)
morphology is small, the effects on the quantitative parameters $C$
and $A$ can be substantial. Half-light radius and surface-brightness
are only weakly affected.

\bigskip

\noindent Fig. 9.--- Resolution dependence of image concentration,
$C$, for the galaxies in the Frei \etal catalog: $\Delta C$ versus the
half-light radius $R_e$ (in pixel units of the block-averaged images).
$\Delta C$ is the difference between the concentration index for a
given simulated value of $R_e$ relative to the original concentration
value (i.e. that value measured on the observed image). Measurements
for six definitions of the concentration index are plotted (two types
per panel, labeled by line-type).  The central line (bold) is the
median value of this difference, and the bounding lines are the 25$\%$
and 75$\%$ values, i.e. 50$\%$ of the simulations are contained
between the upper and lower lines for each index.

\bigskip

\noindent Fig. 10.--- Wavelength dependence of structural parameters
for galaxies in the Frei \etal sample, plotted versus galaxy
rest-frame color. Dashed lines show the mean
differences between blue and red bands and the error bars show the
1$\sigma$ dispersions for three bins in color: $\bv < 0.62$
(late-type), $0.62<\bv<0.87$ (intermediate-type), and $\bv>0.87$
(early-type). Top panel: Image concentration $C_B-C_R$. Nearly all
galaxies are more highly concentrated in the red band than in the
blue, and thus fall below the dotted line at $C_B-C_R=0$.  This
difference is slightly larger for galaxies with intermediate-type
morphology. Middle panel: Image asymmetry $A_B-A_R$. Late- and
intermediate-type galaxies are more asymmetric in the blue band than
in the red band. Red objects are generally very symmetric in both
bands, and have $A_B-A_R \sim 0$. This panel can be compared to Figure
2 in Conselice \etal (1999) where $A_B-A_R$ is plotted versus $A_R$.
Since asymmetry and color are strongly correlated for the Frei \etal
sample (as seen in Figure 3), the trend in Conselice's plot is
similar to what is shown here. Bottom panel: Half-light radius
$R_{e,B} - R_{e,R}$. Although the scatter in this diagram is
relatively large, it is clear that the half-light radius shows little
wavelength dependence over this wavelength range (cf. de Jong
1995). Objects of intermediate \bv \ color tend to be slightly larger
in the blue band than in the red, but this trend is not seen for
either the bluest or the reddest objects.
 
\bigskip

\noindent Fig. 11.--- Axis ratio dependence of image concentration $C$
for galaxies of different morphological types in the Frei \etal
sample. The dotted line separates the sample into two bins at
log$_{10} [a/b]=0.3$, corresponding to an inclination of
60\arcdeg. The dashed lines and error bars show the mean and 1$\sigma$
dispersion for each morphological type and bin. Labeled objects are
discussed in the text.  Top panel: Late-type objects with high
inclination have slightly higher measured $C$ than more face-on
objects. Middle panel: For intermediate-type galaxies the measured
concentration indices show no correlation with the axial ratio
$[a/b]$. Bottom panel: For early-type galaxies, a tendency can be seen
where objects with larger axial ratio $[a/b]$ are measured to have
lower image concentration.

\bigskip

\noindent Fig. 12.--- Axis ratio dependence of half-light radius $R_e$
for galaxies of different morphological types in the Frei \etal
sample.  The dotted line separates the sample into two bins at
log$_{10} [a/b]=0.3$, corresponding to an inclination of
60\arcdeg. The dashed lines and error bars show the mean and 1$\sigma$
dispersion for each morphological type and bin. Top panel: The
measured $R_e$ are slightly larger for late-type objects with high
axial ratios.  Middle panel: Intermediate-type objects have somewhat
smaller $R_e$ for high values of $[a/b]$. In both of these panels the
scatter is large and the differences between the bins are small.
Bottom panel: Early-type galaxies with larger axial ratio $[a/b]$ are
measured to have $\lesssim 30\%$ smaller half-light radii.

\input bershady_figures.tex

\end{document}

%% file: bershady_tables.tex
\clearpage

\begin{figure}
\plotfiddle{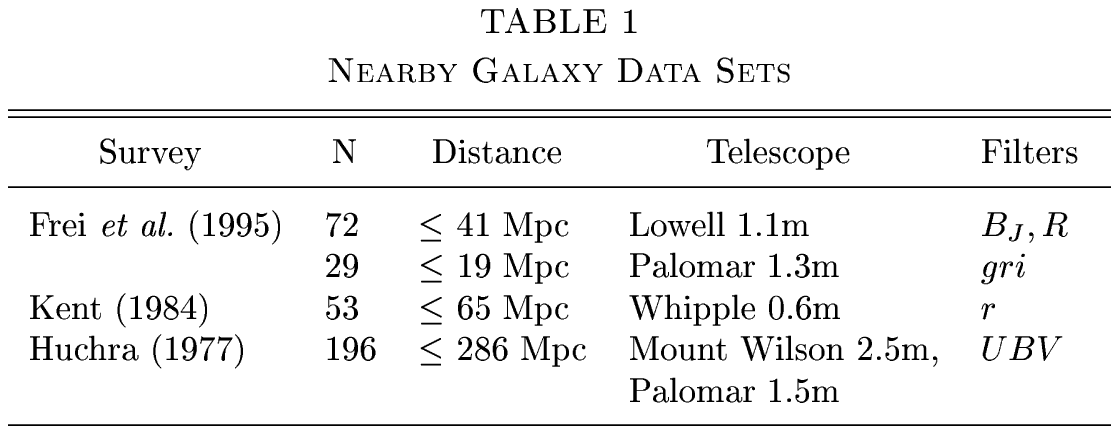}{8in}{0}{100}{100}{-300}{-100}
\end{figure}

\clearpage

\begin{figure}
\plotfiddle{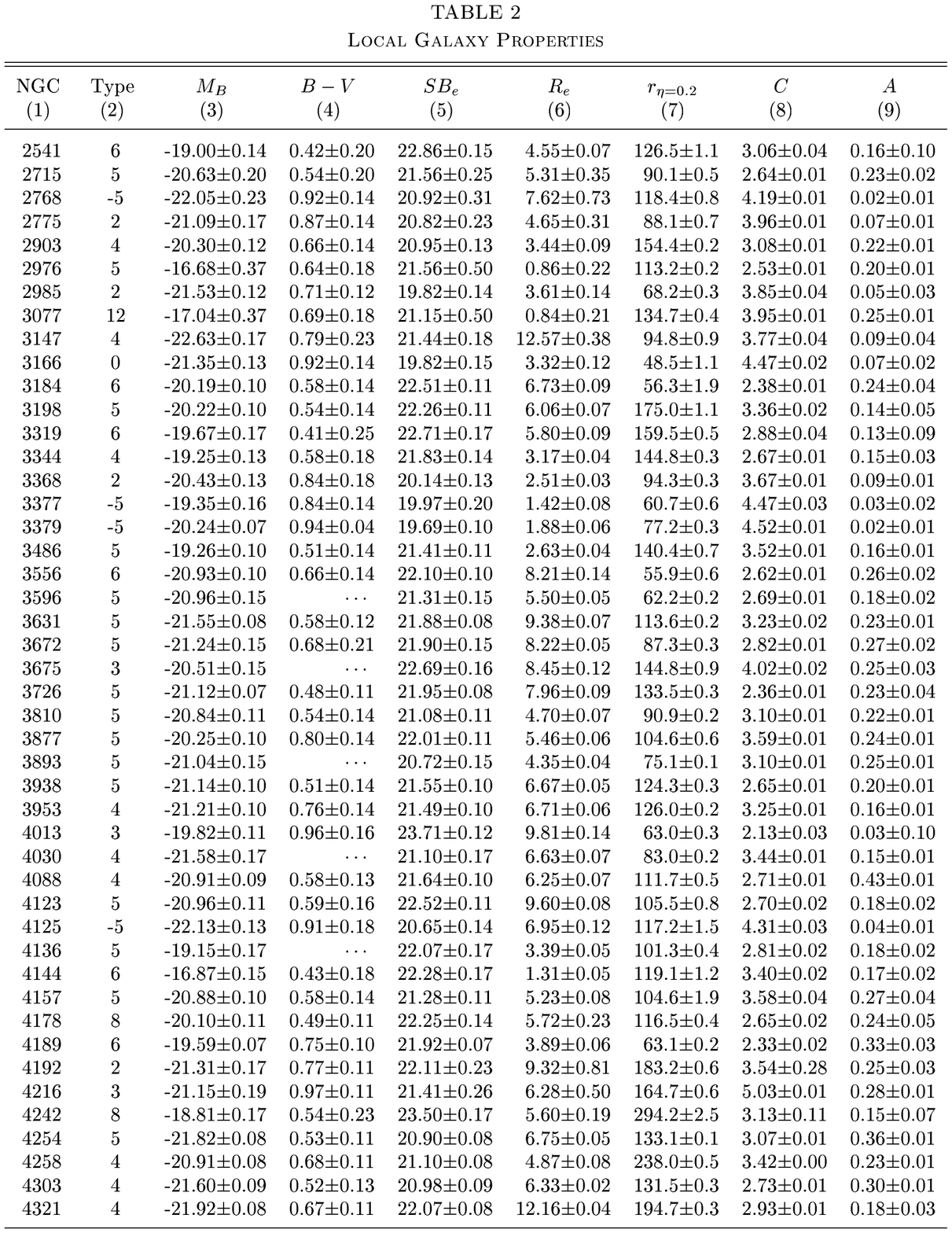}{8in}{0}{100}{100}{-300}{-100}
\end{figure}

\clearpage

\begin{figure}
\plotfiddle{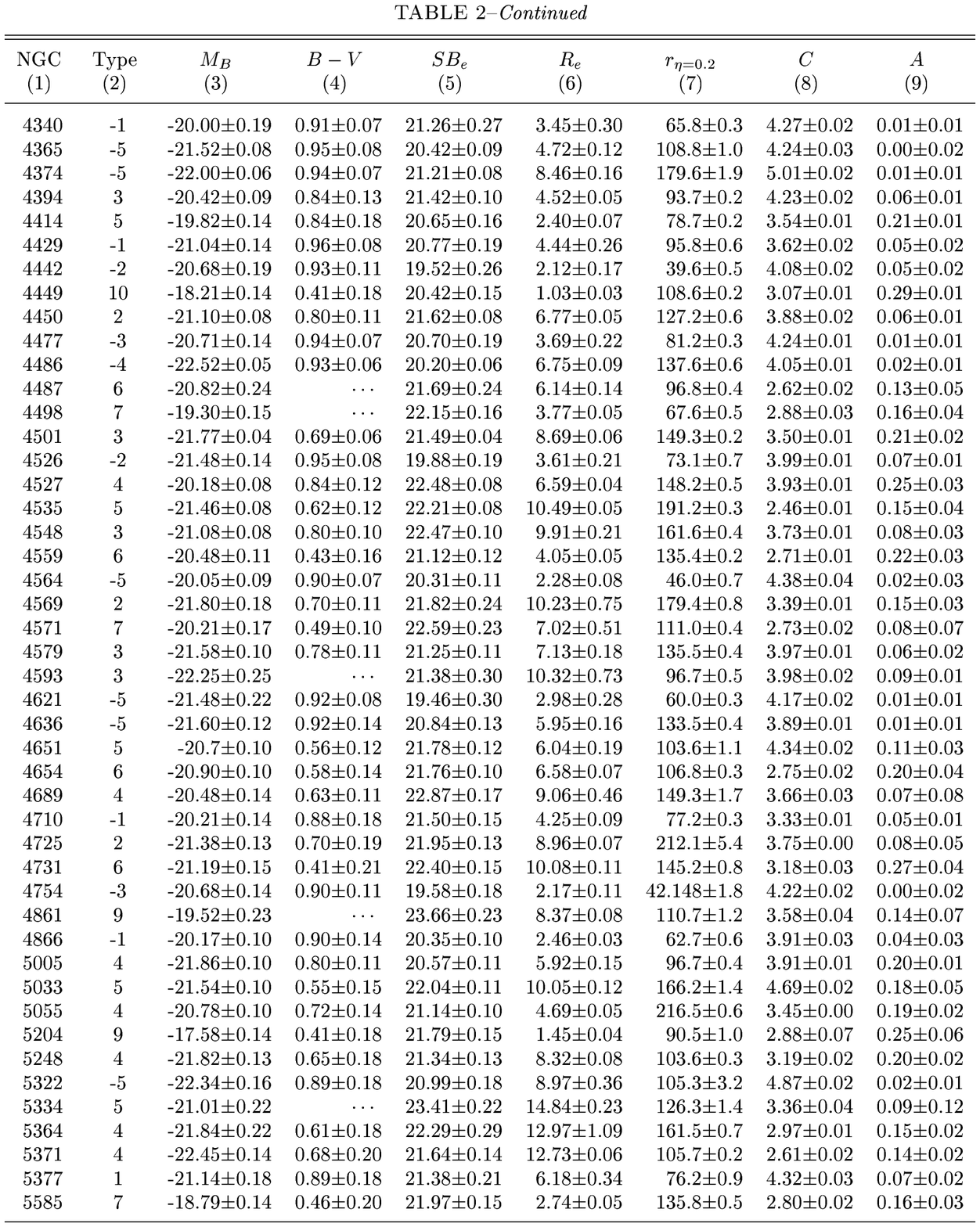}{8in}{0}{100}{100}{-300}{-100}
\end{figure}

\clearpage

\begin{figure}
\plotfiddle{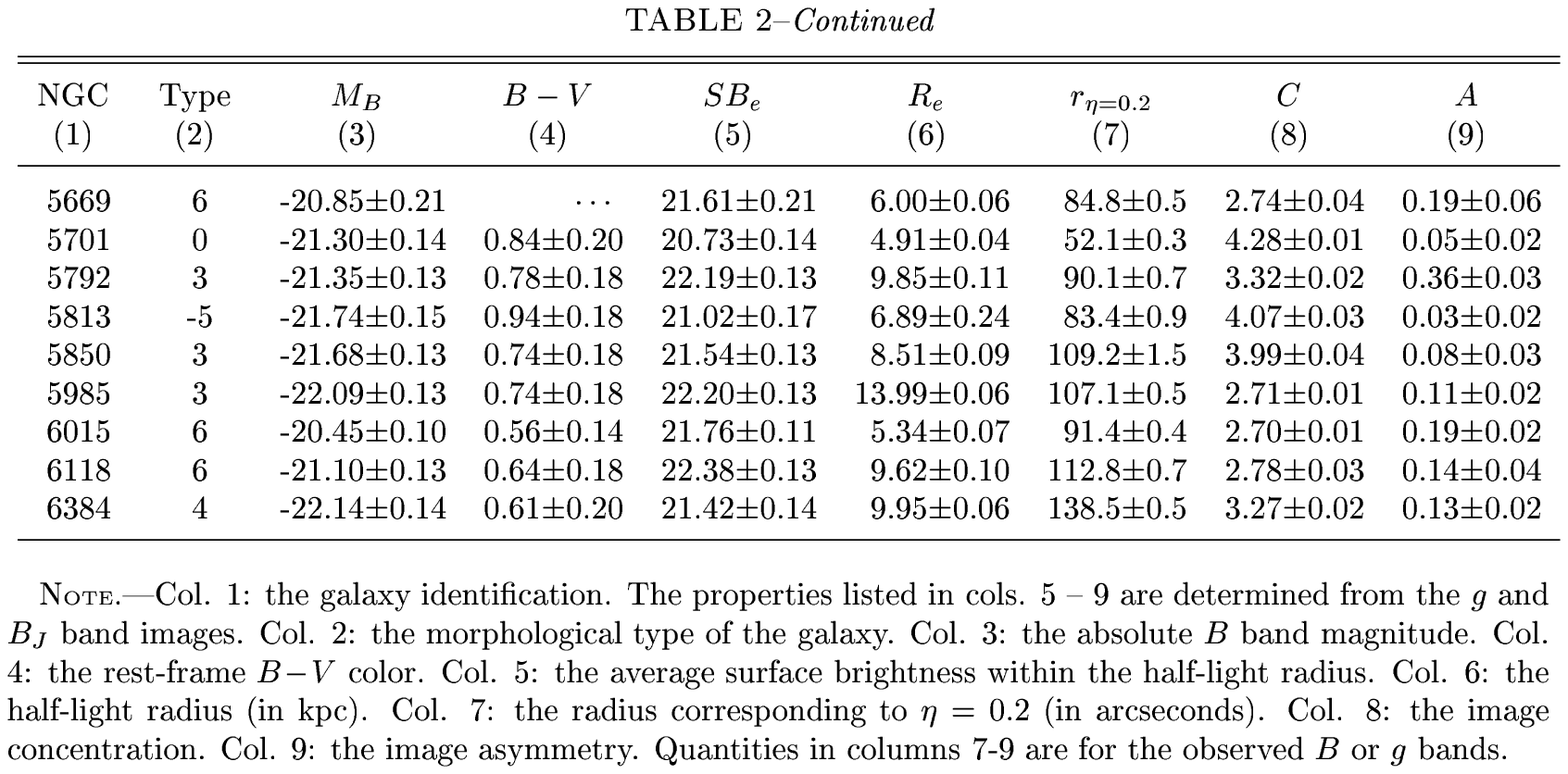}{8in}{0}{100}{100}{-300}{-100}
\end{figure}

\clearpage

\begin{figure}
\plotfiddle{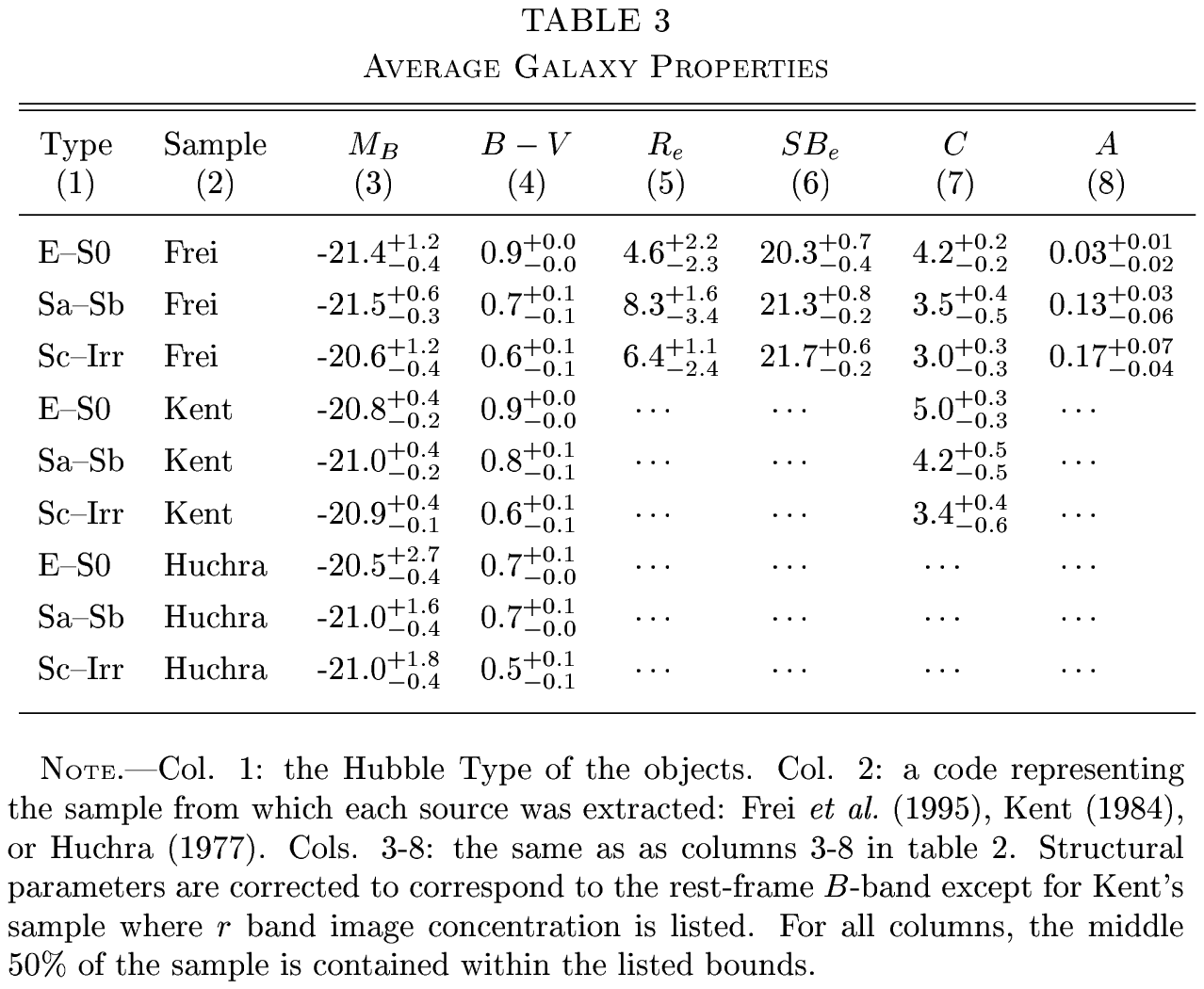}{8in}{0}{100}{100}{-300}{-100}
\end{figure}

\clearpage

\begin{figure}
\plotfiddle{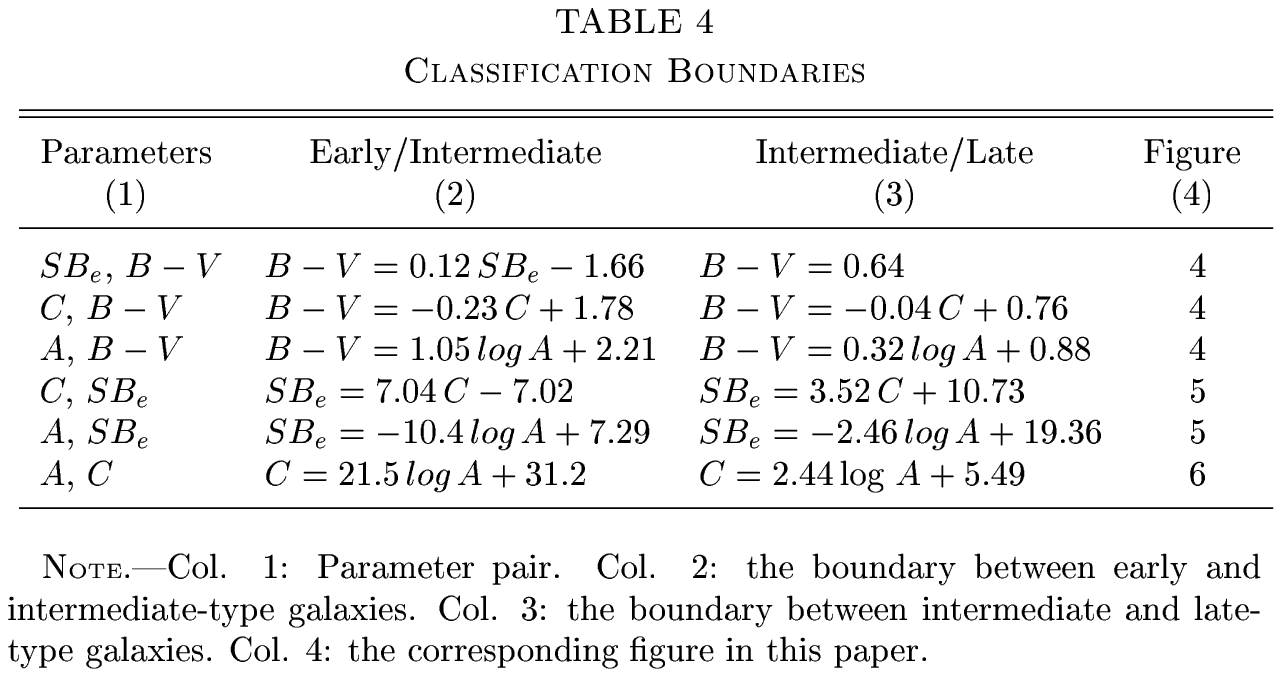}{8in}{0}{100}{100}{-300}{-100}
\end{figure}

\clearpage

%% file: bershady_figures.tex
\clearpage

\begin{figure}
\plotfiddle{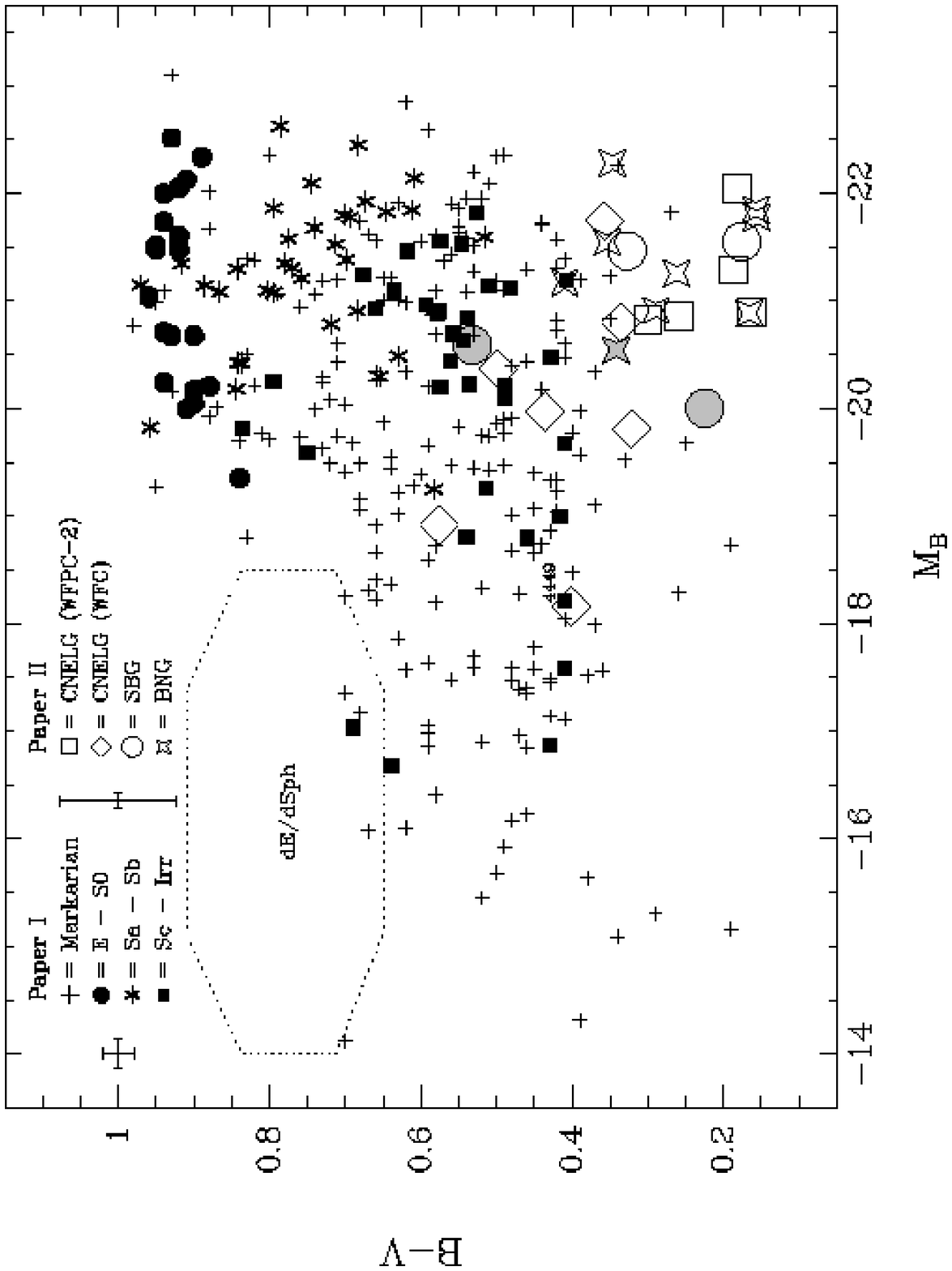}{1in}{-90}{75}{75}{-310}{250}
\vskip 3.0in
\caption{}\label{f1}
\end{figure}

\clearpage

\begin{figure}
\plotfiddle{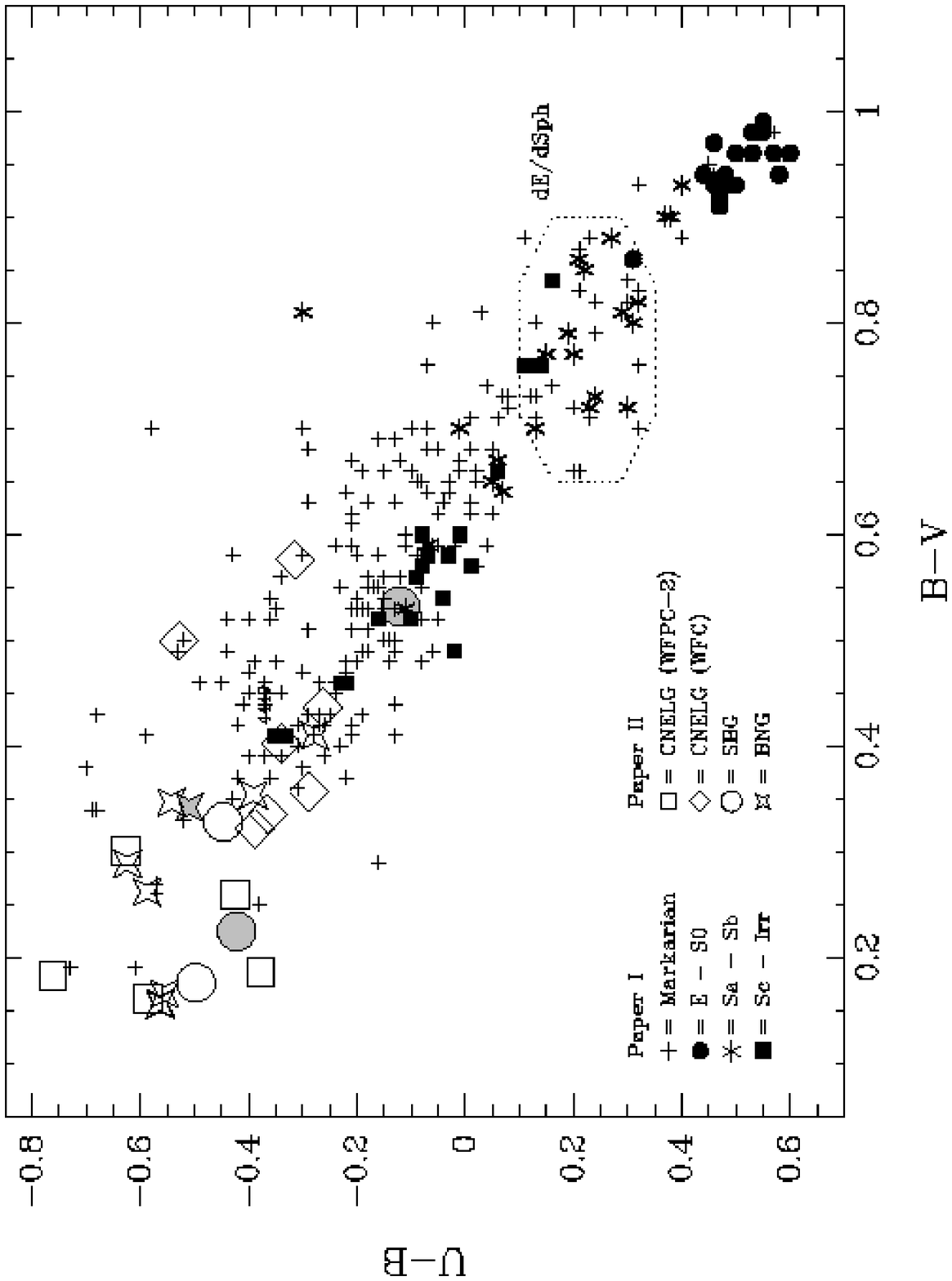}{1in}{-90}{75}{75}{-310}{250}
\vskip 3.0in
\caption{}\label{f2}
\end{figure}

\clearpage

\begin{figure}
\plotfiddle{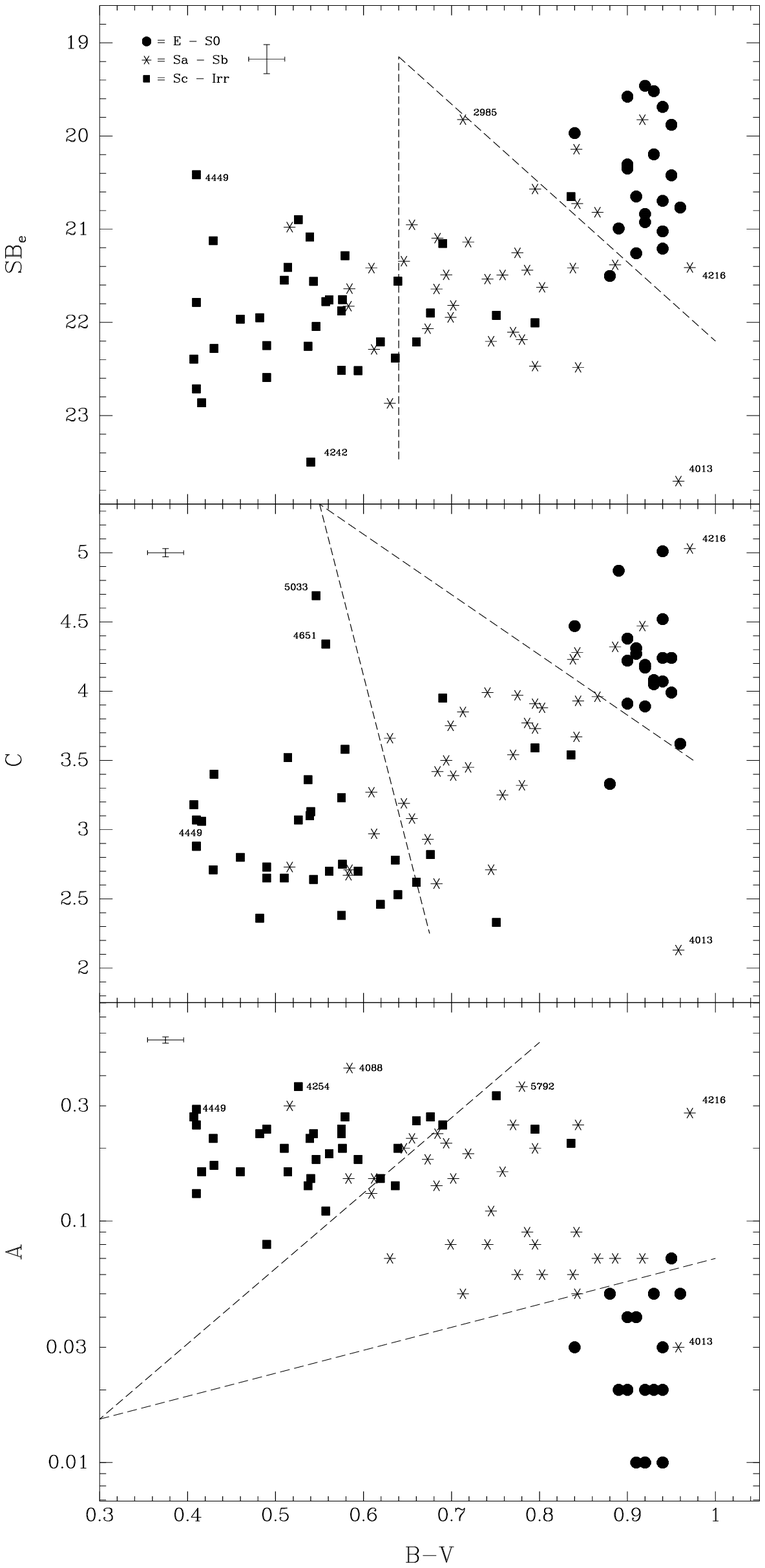}{7in}{0}{90}{90}{-270}{-150}
\vskip 1.0in
\caption{}\label{f4}
\end{figure}

\clearpage

\begin{figure}
\plotfiddle{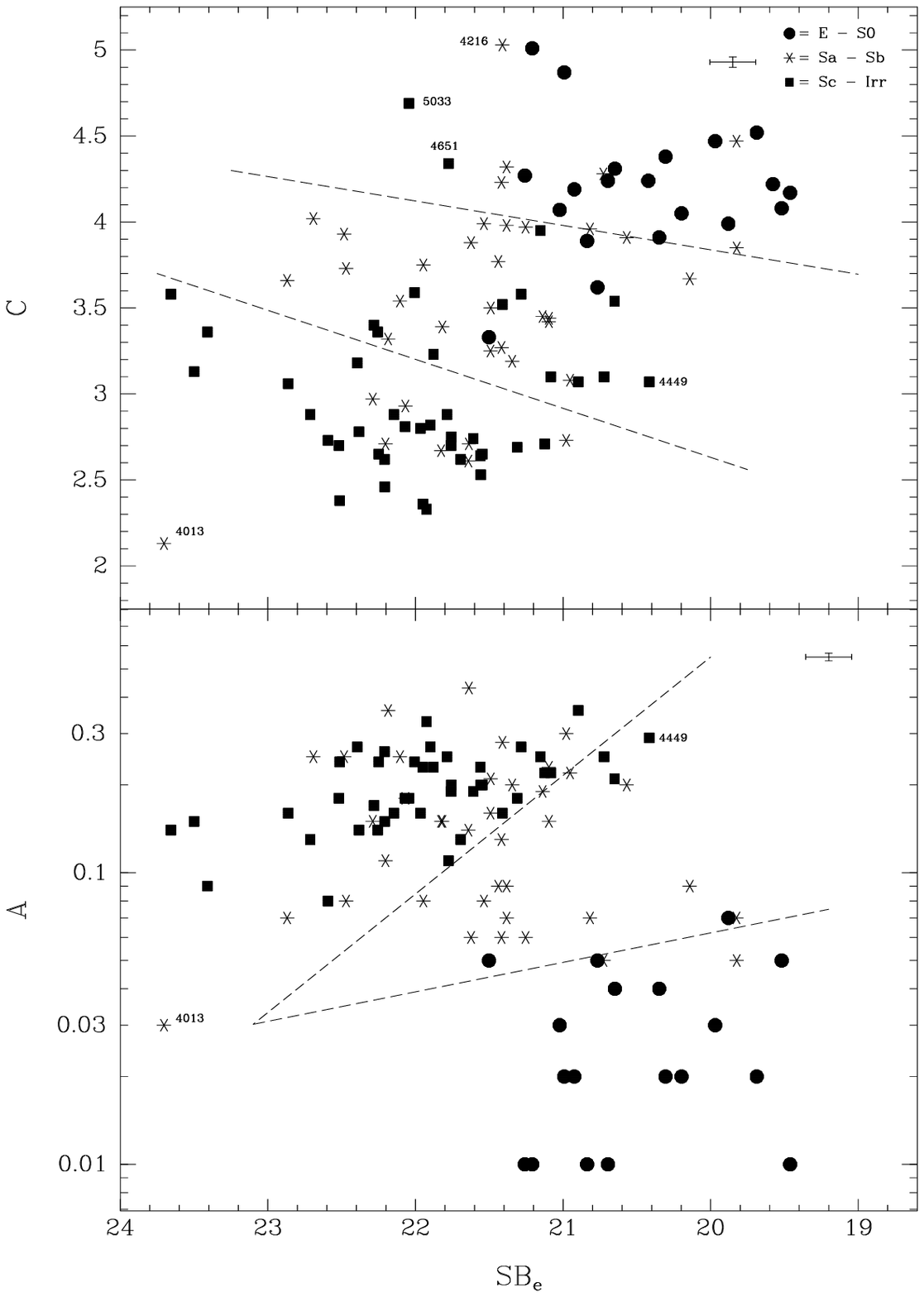}{7in}{0}{90}{90}{-270}{-150}
\vskip 1.0in
\caption{}\label{f5}
\end{figure}

\clearpage

\begin{figure}
\plotfiddle{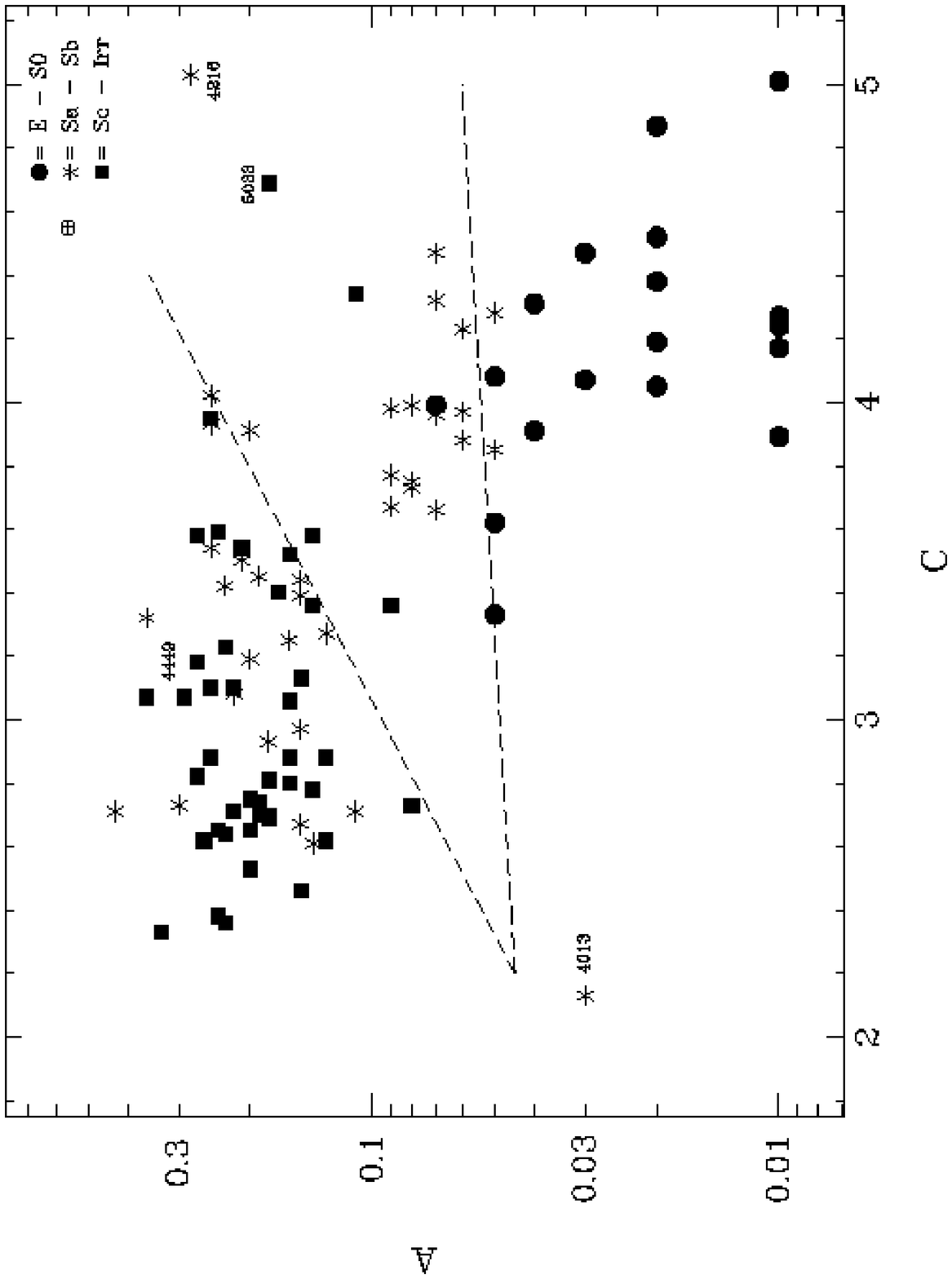}{1in}{-90}{75}{75}{-310}{250}
\vskip 3.0in
\caption{}\label{f6}
\end{figure}

\clearpage

\begin{figure}
\plotfiddle{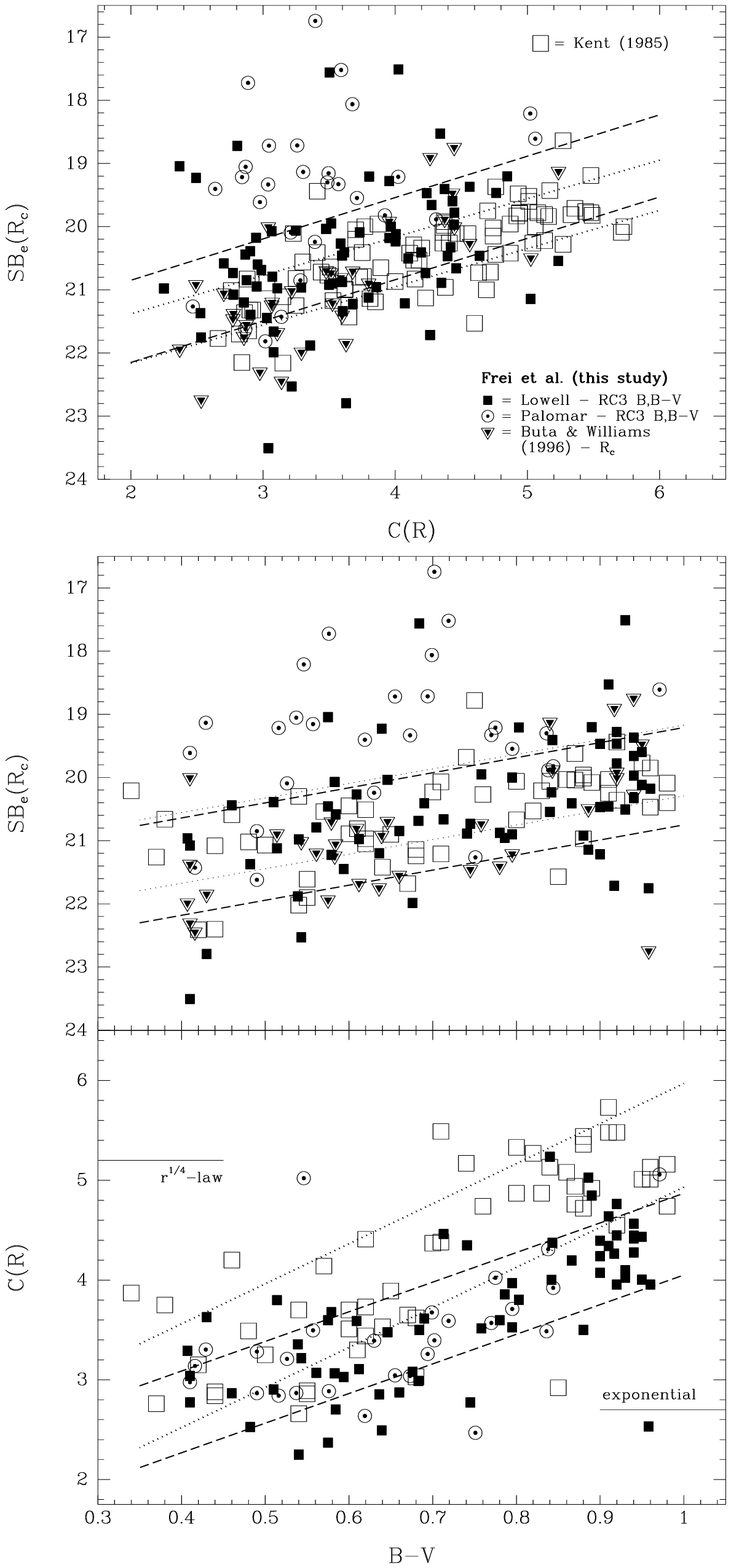}{7in}{0}{90}{90}{-270}{-150}
\vskip 1.0in
\caption{}\label{f7}
\end{figure}

\clearpage

\begin{figure}
\plotfiddle{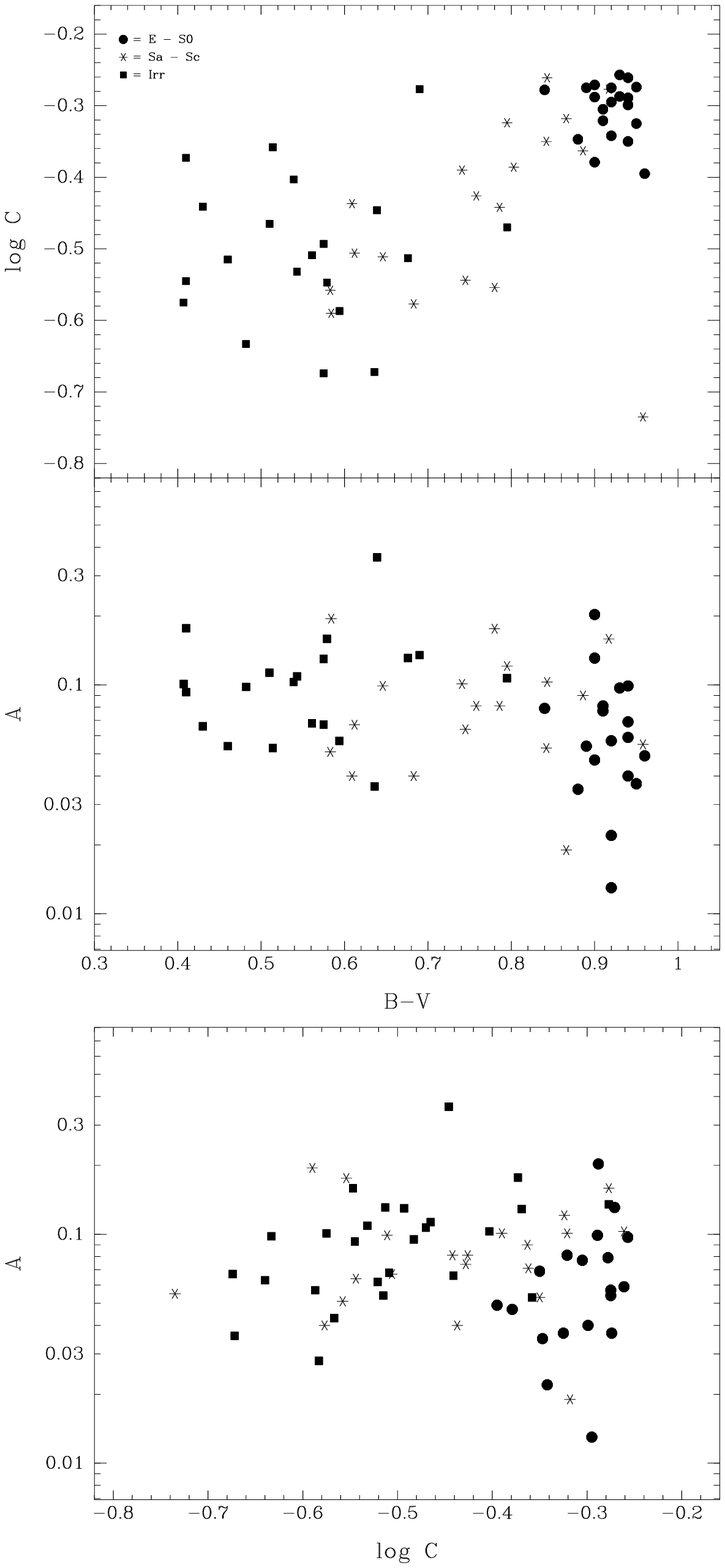}{7in}{0}{90}{90}{-270}{-150}
\vskip 1.0in
\caption{}\label{f8}
\end{figure}

\clearpage

\begin{figure}
\plotfiddle{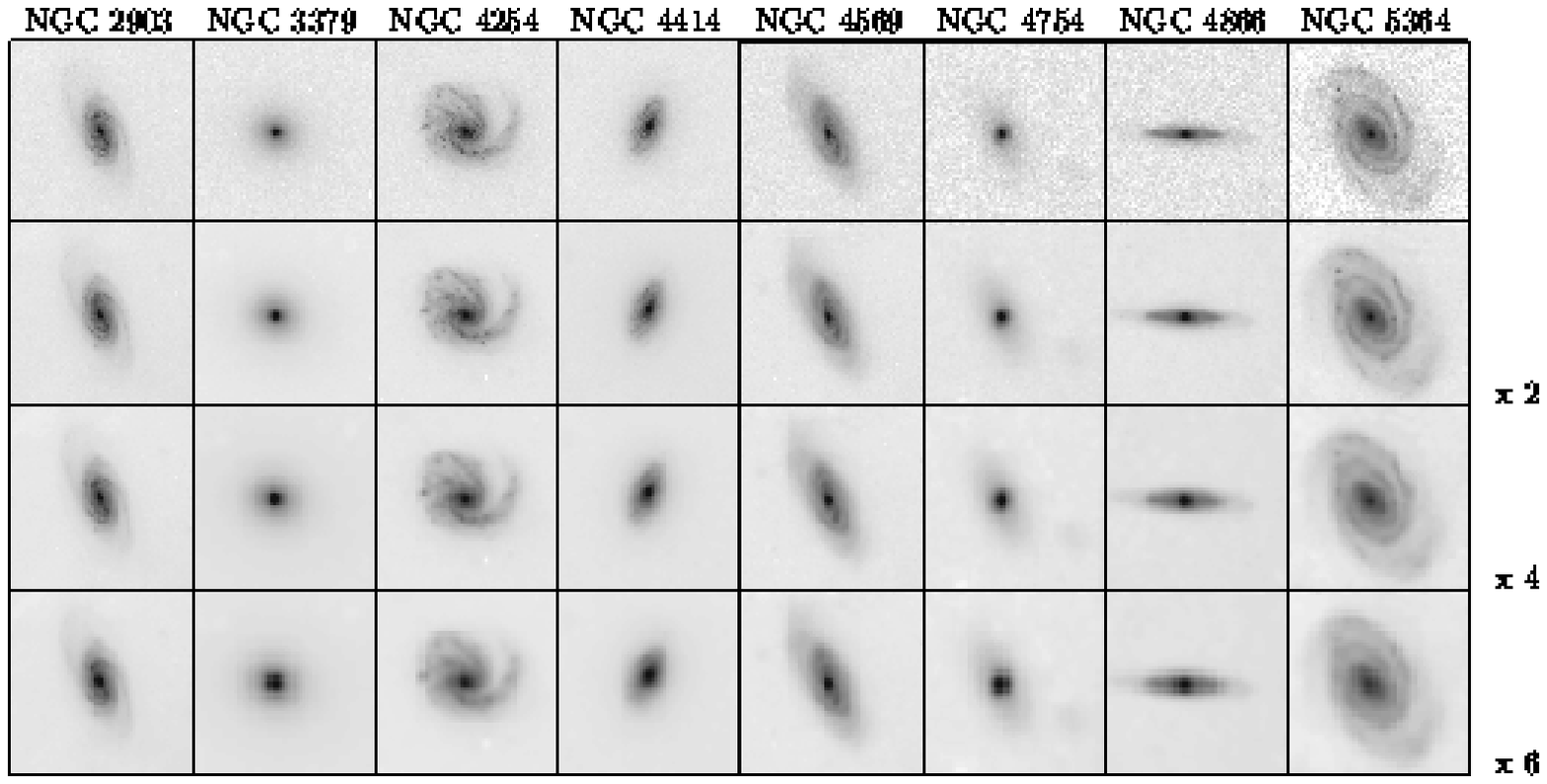}{1in}{0}{100}{100}{-300}{-400}
\vskip 3.0in
\caption{}\label{f9}
\end{figure}

\clearpage

\begin{figure}
\plotfiddle{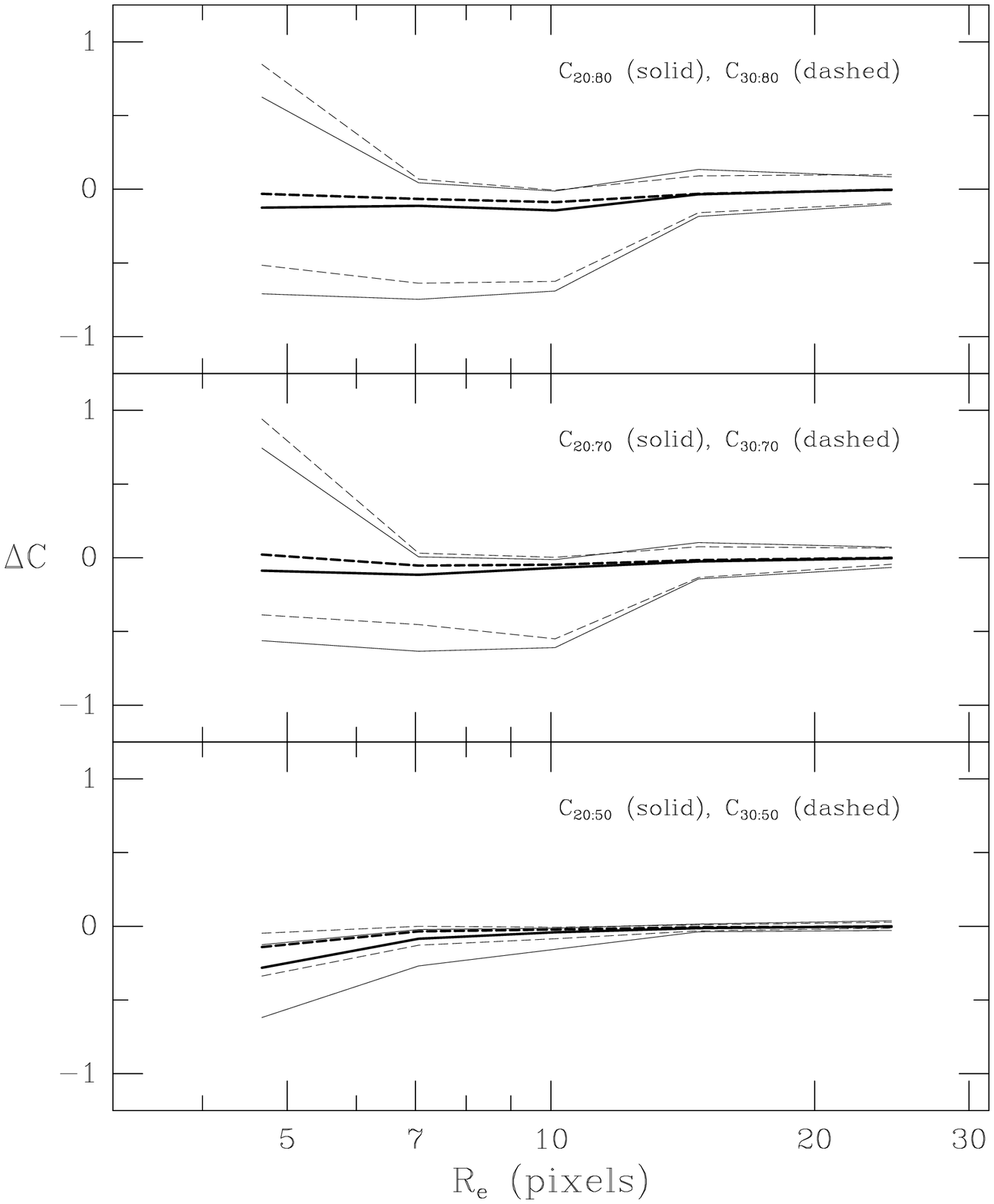}{7in}{0}{90}{90}{-270}{-150}
\vskip 1.0in
\caption{}\label{f10}
\end{figure}

\clearpage

\begin{figure}
\plotfiddle{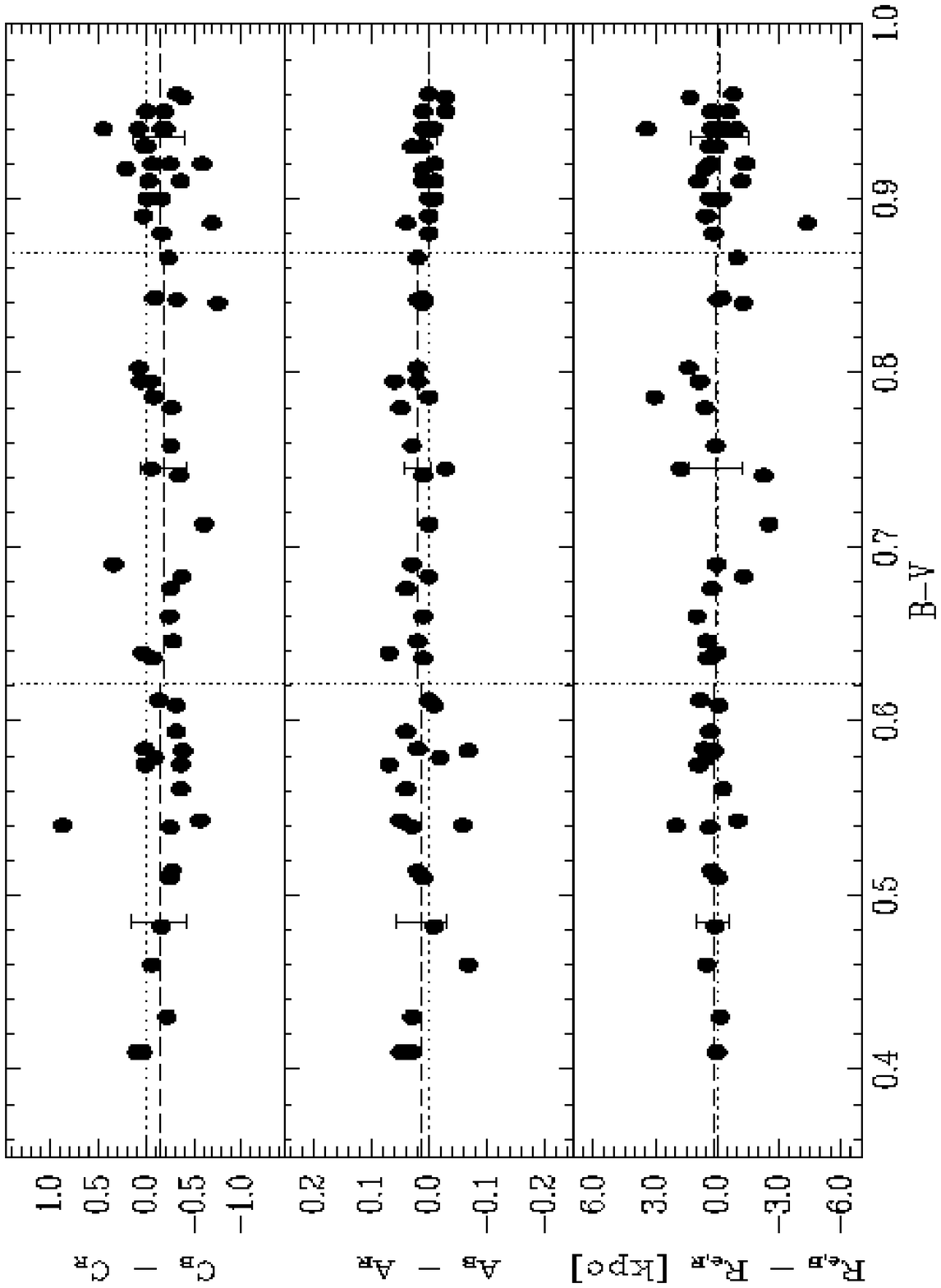}{1in}{-90}{75}{75}{-290}{250}
\vskip 3.0in
\caption{}\label{f11}
\end{figure}

\clearpage

\begin{figure}
\plotfiddle{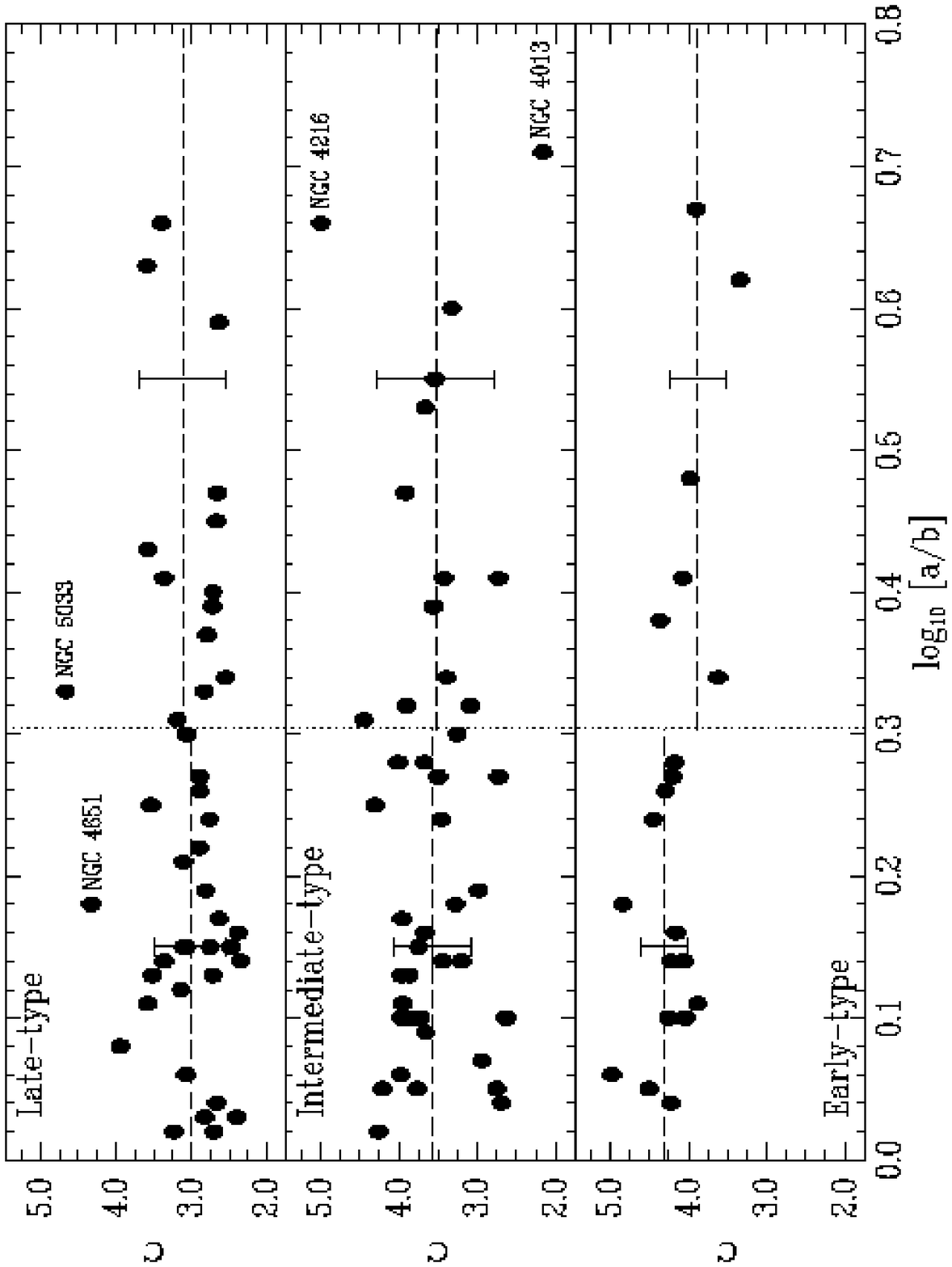}{1in}{-90}{75}{75}{-290}{250}
\vskip 3.0in
\caption{}\label{f12}
\end{figure}

\clearpage

\begin{figure}
\plotfiddle{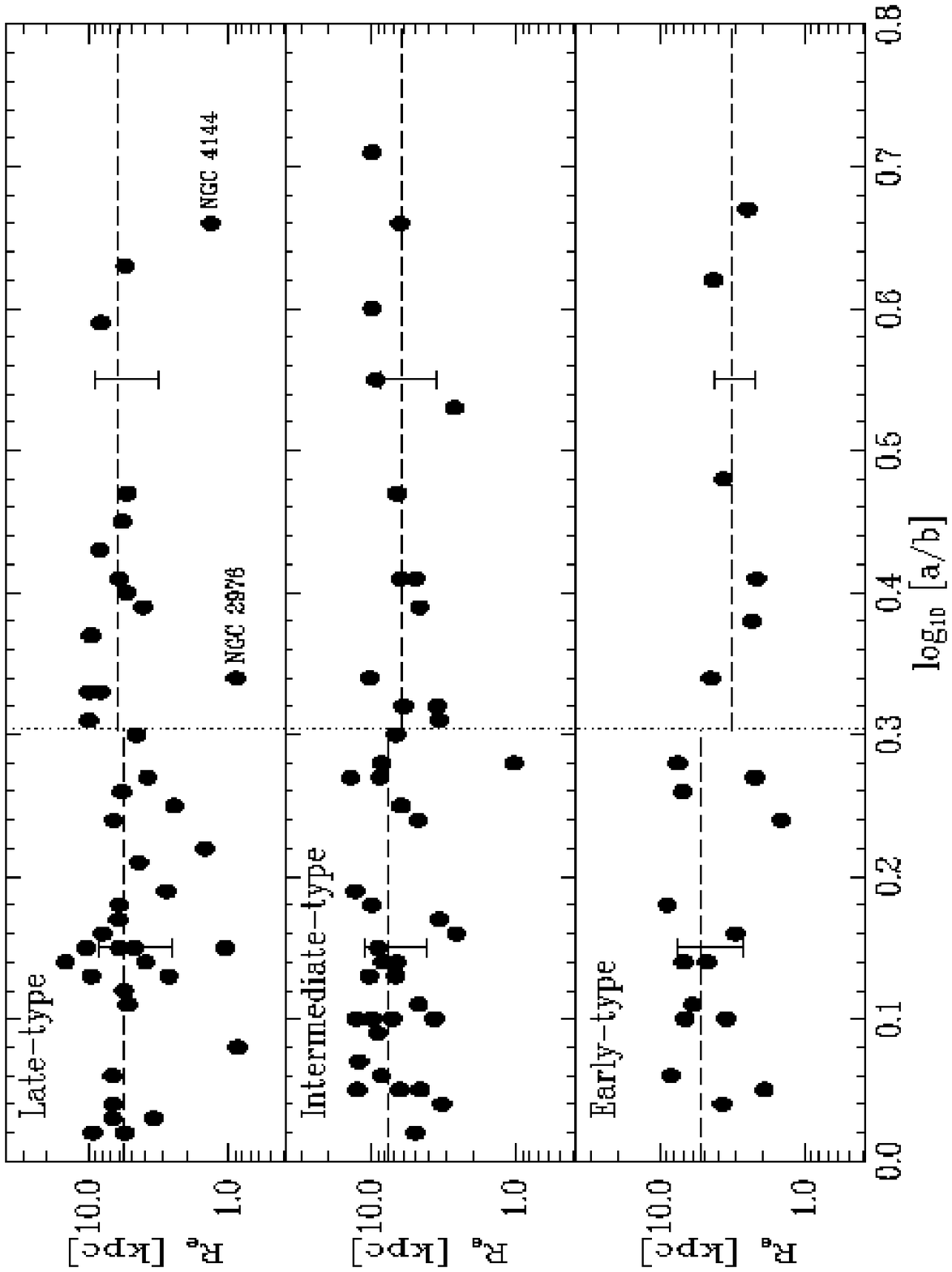}{1in}{-90}{75}{75}{-290}{250}
\vskip 3.0in
\caption{}\label{f13}
\end{figure}